\documentclass[12pt]{article}
\usepackage{amssymb,amsmath,epsfig}
\usepackage{graphicx}\allowdisplaybreaks
\begin{document}

\title{\bf Compact Objects by Gravitational Decoupling in $f(R)$ Gravity}

\author{M. Sharif \thanks{msharif.math@pu.edu.pk} and Mariyah Aslam
\thanks{mariyahaslam230@gmail.com}\\
Department of Mathematics, University of the Punjab,\\
Quaid-e-Azam Campus, Lahore-54590, Pakistan.}

\date{}

\maketitle
\begin{abstract}
The objective of this paper is to discuss anisotropic solutions
representing static spherical self-gravitating systems in $f(R)$
theory. We employ the extended gravitational decoupling approach and
transform temporal as well as radial metric potentials which
decomposes the system of non-linear field equations into two arrays:
one set corresponding to seed source and the other one involves
additional source terms. The domain of the isotropic solution is
extended in the background of $f(R)$ Starobinsky model by employing
the metric potentials of Krori-Barua spacetime. We determine two
anisotropic solutions by employing some physical constraints on the
extra source. The values of unknown constants are computed by
matching the interior and exterior spacetimes. We inspect the
physical viability, equilibrium and stability of the obtained
solutions corresponding to the star Her X-I. It is observed that one
of the two extensions satisfies all the necessary physical
requirements for particular values of the decoupling parameter.
\end{abstract}

\section{Introduction}

The study of the vast universe offers insights into its origin and
puzzling nature. In the present era, different astrophysical
phenomena such as the formation and evolution of cosmic structures
have captured the attention of many researchers. Among the cosmic
entities, stars are considered as the elementary constituents of
galaxies which are organized systematically in a cosmic web. The
collapse of stars due to the inward pull of gravity results in the
formation of new compact objects. In order to explore the interior
geometry of these objects, we need analytical solutions of the
non-linear field equations. Despite the non-linearity of these
partial differential equations, many researchers have constructed
exact viable astrophysical and cosmological solutions. Schwarzschild
\cite{1}, developed the first solution of Einstein filed equations
for an isotropic sphere in the vacuum.

It has been observed that the presence of interacting nuclear matter
in dense celestial objects leads to the generation of anisotropy
\cite{2}. Fluid configurations with condensed pion like neutron
stars are also anisotropic in nature \cite{3}. The impact of
pressure anisotropy on features of stellar structures is apparent in
various studies of charged or uncharged compact objects. In 1974,
the effects of anisotropy on relativistic spherical objects were
studied by using specific equations of state (EoS) and an increase
in redshift was noted in static models with particular forms of
anisotropy \cite{4}. Santos and Herrera \cite{5} examined the origin
of anisotropy in general relativity (GR) and studied its impact on
the stability of self-gravitating systems. Harko and Mak \cite{6}
developed well-behaved anisotropic spherical solutions and examined
their physical properties. In 2002, Dev and Gleiser \cite{7}
discussed the factors contributing to pressure anisotropy in stellar
objects. Hossein et al. \cite{8} constructed anisotropic models for
different values of the cosmological constant and used cracking
approach to check their stability. Paul and Deb \cite{9} examined
feasible anisotropic solutions in hydrostatic equilibrium and showed
that the anisotropic stars corresponding to these solutions
represent viable behavior. In 2016, Arba\~{n}il and Malheiro
\cite{10} considered the MIT bag model and discussed the stability
of a strange star comprising of anisotropic fluid. Murad \cite{11}
developed a model of anisotropic strange star by incorporating the
effects of charge for particular forms of radial metric function.

It is a difficult task to extract an anisotropic solution of the
non-linear system of field equations due to the greater number of
unknowns as compared to the number of equations. To overcome this
problem, various techniques have been introduced which aid in the
construction of feasible solutions. In this regard, the
gravitational decoupling technique through minimal geometric
deformation (MGD), proposed by Ovalle \cite{12}, determines new
solutions corresponding to different relativistic distributions in
astrophysics. This approach deforms the radial metric component and
generates two sets of differential equations from the system of
field equations. One system incorporates the seed source while the
other one is governed by the impact of the additional source. Both
sets are solved separately and the solution of the whole system is
obtained by using the superposition principle. The MGD technique
prevents the exchange of energy between matter sources and preserves
the spherical symmetry of the self-gravitating system.

Following the MGD scheme, Ovalle and Linares \cite{13} evaluated a
solution in braneworld and deduced that the compactness factor
reduces due to the bulk effects of fluid distribution. Later, in
2018, Ovalle and his collaborators \cite{14} developed an
anisotropic interior solution for perfect fluid distribution by the
inclusion of an additional gravitational source. Gabbanelli et al.
\cite{15} inspected the salient features of the anisotropic
extension of Durgapal-Fuloria model via gravitational decoupling.
Sharif and Sadiq \cite{16} devised charged anisotropic models
through this scheme and examined the physical features of stellar
bodies. Morales and Tello-Ortiz \cite{17} used Heintzmann solution
as a seed source and examined the static spherical anisotropic model
under the influence of electromagnetic field. Graterol \cite{18}
adopted this approach to extend the domain of isotropic Buchdahl
solution via some physical constraints. Contreras and Bargue\~{n}o
\cite{19} obtained an anisotropic static BTZ model by employing MGD
in $(2+1)$-dimensional spacetime. Estrada and Prado \cite{20}
explored the higher-dimensional extension of MGD approach to
construct well-behaved analytical solutions corresponding to
anisotropic star models. Maurya and Tello-Ortiz \cite{21}
graphically analyzed the physical characteristics of an anisotropic
solution formulated via the MGD approach. Sharif and Ama-Tul-Mughani
\cite{22} discussed the $(2+1)$-dimensional charged string cloud
through this technique. They also formulated analytical solutions of
axially symmetric geometry in the framework of cosmic strings
through gravitational decoupling technique \cite{23*}.

Although the MGD technique has facilitated the study of
self-gravitating objects, it transforms radial coordinate only while
the temporal coordinate remains invariant which gives rise to
certain shortcomings in the decoupling procedure. Since there is no
transfer of energy between matter sources, therefore the interaction
between them is purely gravitational. To resolve these issues,
Casadio et al. \cite{23} proposed an extension of the MGD technique
by implementing radial as well as temporal transformations and
constructed a solution for a static spherical object. However, the
extended method is applicable only in vacuum and fails in the
presence of matter as the conservation law does not hold. Therefore,
the intrinsic features of astrophysical systems cannot be examined
via this approach. Later, in 2019, Ovalle \cite{24} introduced a
novel extension of the MGD approach known as extended geometric
deformation (EGD). He successfully decoupled two static spherically
symmetric gravitational sources and examined its efficiency by
recreating the Reissner-Nordstr\"{o}m solution. Contreras and
Bargue\~{n}o \cite{25} used this technique in $(2+1)$-dimensional
gravity and obtained exterior charged BTZ solution from its vacuum
counterpart. Sharif and Ama-Tul-Mughani employed EGD approach to
compute anisotropic solutions corresponding to Tolman IV \cite{26}
and Krori-Barua \cite{26a} solutions.

Recent study of the universe suggests that strange dark energy is
causing cosmic expansion. The modification of GR helps us to
investigate the mysterious nature of this repulsive energy. The
$f(R)$ theory is one of the simplest modification as it generalizes
GR by replacing Ricci scalar $(R)$ with its generic function in the
action integral \cite{27}. Starobinsky \cite{28} introduced a model
of higher curvature terms $(R+\sigma R^{2})$ to study the
inflationary epoch. Many astrophysicists have discussed other forms
of $f(R)$ to resolve different cosmic problems such as accelerated
cosmic expansion \cite{29} and history of the universe \cite{30}. In
the last decade, numerous work has been done on the viability and
dynamical stability of astrophysical objects in $f(R)$ gravity.
Effects of this theory on the dynamical instability of expansion
free fluid were investigated \cite{33}. Later, Capozziello et al.
\cite{34} discussed the hydrostatic equilibrium and dynamics of
collisionless self-gravitating systems. Researchers have extensively
discussed the collapsing behavior of neutron stars in $f(R)$ theory
\cite{35}. In 2015, the dynamics of a static spherically symmetric
object was investigated by using the Tolman-Oppenheimer-Volkoff
equation \cite{40}. Zubair and Abbas \cite{41} explored the
stability and dynamics of anisotropic compact stars in $f(R)$
background. Recently, researchers have adopted various EoS to
describe the mechanism and salient features of different compact
anisotropic spheres \cite{41**}.

In 2019, Sharif and Waseem \cite{43} used the isotropic Krori-Barua
model for both charged and uncharged spherically symmetric systems
in the $f(R)$ framework to construct viable and stable anisotropic
solutions via MGD. Recently, EGD and MGD approaches have extensively
been used in other modified theories as well \cite{44}. This paper
explores the efficiency of the EGD method in the framework of $f(R)$
gravity. We consider the Krori-Barua solution as a seed source to
construct a new anisotropic solution and inspect salient features of
the extended version. The paper is arranged in the following format.
The next section provides the $f(R)$ field equations with an
additional gravitational source. The gravitational decoupling of
$f(R)$ field equations via EGD technique is presented in section
\textbf{3}. In section \textbf{4}, the junction conditions are
computed by matching the interior with exterior Schwarzschild
solution. In section \textbf{5}, we construct two anisotropic static
models by applying physical constraints on the additional
gravitational source and analyze the validity of both solutions.
Finally, section \textbf{6} summarizes the obtained results.

\section{Field Equations in $f(R)$ Gravity}

The modified Einstein-Hilbert action for $f(R)$ gravity has the form
\cite{47}
\begin{equation}\label{1}
\mathcal{I}_{f(R)}=\int\sqrt{-g}\left[\frac{f(R)}{2
\kappa}+\mathcal{L}_m+\chi\mathcal{L}_{\Theta}\right]d^4x,
\end{equation}
where $g$ and $f(R)$ represent determinant of the metric tensor and
arbitrary function of Ricci scalar, respectively. Also, $\kappa=1$
(in relativistic units) represents the coupling constant, while
$\chi$ symbolizes the decoupling parameter. Furthermore,
$\mathcal{L}_{m}$ and $~\mathcal{L}_{\Theta}$ are the Lagrangian
densities for seed source and additional source, respectively. The
field equation obtained by varying Eq.(\ref{1}) with respect to the
metric is given as
\begin{equation}\label{2}
f_RR_{\xi\eta}-\frac{1}{2}g_{\xi\eta}f(R)-(\nabla_\xi\nabla_\eta
-g_{\xi\eta}\Box)f_R=T^{(m)}_{\xi\eta},
\end{equation}
where $f_{R}=\frac{\partial f}{\partial R}$, ~$\Box$ is the
d'Alembertian operator defined as
$\Box=g^{\xi\eta}\nabla_\xi\nabla_\eta$,~ $\nabla_\xi$ represents
the covariant derivative and $T^{(m)}_{\xi\eta}$ is the standard
energy-momentum tensor for perfect fluid given as
\begin{equation}\label{3}
T^{(m)}_{\xi\eta}=(\rho+p)u_{\xi}u_{\eta}-pg_{\xi\eta}.
\end{equation}
Here, $\rho,~p$ and $u_{\xi}$ denote the energy density, pressure
and four velocity of the fluid, respectively. An alternative form of
Eq.(\ref{2}) is
\begin{equation}\label{4}
R^{\xi}_{\eta}-\frac{1}{2}R\delta^{\xi}_{\eta}=\frac{1}{f_R}T^{\xi(tot)}_{\eta},
\end{equation}
where $T^{\xi(tot)}_{\eta}$ is the energy-momentum tensor describing
the internal configuration of the stellar object and is given as
\begin{equation}\label{5}
T^{\xi(tot)}_{\eta}=T^{\xi(m)}_{\eta}+\chi\Theta^{\xi}_{\eta}+T^{\xi(F)}_{\eta}
={\widetilde{T}}^{\xi}_{\eta}+\chi\Theta^{\xi}_{\eta},
\end{equation}
where
$\widetilde{T}^{\xi}_{\eta}=T^{\xi(m)}_{\eta}+T^{\xi(F)}_{\eta}$ and
$T^{\xi(F)}_{\eta}=\left(\frac{f(R)-Rf_R}{2}\right)\delta^{\xi}_{\eta}
+(\nabla^\xi\nabla_\eta-\delta^{\xi}_{\eta}\Box)f_R.$ Moreover,
$\Theta^{\xi}_{\eta}$ represents the additional source which is
coupled to gravity through a free parameter $\chi$. This source term
comprises of new fields which induce anisotropy in self-gravitating
bodies.

The interior line element describing the spherical structure of a
static spacetime has the form
\begin{equation}\label{7}
ds^2_{-}=e^{\mu(r)}dt^2-e^{\lambda(r)}dr^2-r^2(d\theta^2+\sin^2\theta
d\phi^2),
\end{equation}
where $\mu(r)$ and $\lambda(r)$ are unknown metric potentials. Here,
subscript ``$-$'' represents the interior spacetime. The $f(R)$
field equations corresponding to Eq.(\ref{4}) turn out to be
\begin{eqnarray}\nonumber
\rho+\chi\Theta^{0}_{0}&=&e^{-\lambda}\left[\left(
\frac{\mu''}{2}+\frac{\mu'}{r}+\frac{\mu^{'2}}{4}
-\frac{\mu'\lambda'}{4}\right)f_{R}-\left(\frac{2}{r}
-\frac{\lambda'}{2}\right)f_{R}'-f_{R}''\right]\\\label{8}
&-&\frac{f}{2},\\\nonumber
p-\chi\Theta^{1}_{1}&=&e^{-\lambda}\left[\left(
\frac{\lambda'}{r}-\frac{\mu^{'2}}{4}-\frac{\mu''}{2}
+\frac{\mu'\lambda'}{4}\right)f_{R}+\left(\frac{\mu'}{2}
+\frac{2}{r}\right)f_{R}'\right]\\\label{9}
&+&\frac{f}{2},\\\nonumber
p-\chi\Theta^{2}_{2}&=&e^{-\lambda}\left[\left(
\frac{\lambda'}{2r}-\frac{\mu'}{2r}-\frac{1}{r^2}
+\frac{e^\lambda}{r^2}\right)f_{R}-\left(\frac{\lambda'
-\mu'}{2}-\frac{1}{r}\right)f_{R}'\right.\\\label{10}
&+&\left.f_{R}''\right]+\frac{f}{2},
\end{eqnarray}
where prime denotes derivative with respect to $r$. In $f(R)$
theory, the conservation of the considered setup is expressed as
\begin{equation}\label{11}
p'+\frac{\mu'}{2}(\rho+p)-\chi\left[(\Theta^{1}_{1})'
+\frac{\mu'}{2}(\Theta^{1}_{1}-\Theta^{0}_{0})+\frac{2}{r}(\Theta^{1}_{1}-\Theta^{2}_{2})\right]=0.
\end{equation}
We can regain the conservation equation for the perfect fluid by
setting $\chi=0$.

Scalar fields consistent with theories of superstring and
supergravity have been utilized to formulate several inflationary
models representing the primordial universe. The inflationary model
suggested by Starobinsky \cite{28} is given as
\begin{equation}\label{12}
f(R)=R+\sigma R^2,
\end{equation}
where $\sigma$ is a constant ($\sigma>0$) and $f_{RR}>0$. In this
model, the term $\sigma R^{2}$ describes the exponential expansion
of the universe. Moreover, this model is consistent with the
anisotropic temperature detected in Cosmic Microwave Background.
Thus, it can be used as a reliable alternative for the inflationary
models \cite{51}. Researchers have determined that the value of
$\sigma$ corresponding to celestial objects lies between $0$ and $6$
\cite{41}. It is worth mentioning here that the results of GR can be
recovered for $\sigma=0$.

The field equations corresponding to Eq.(\ref{12}) are expressed as
\begin{eqnarray}\label{13}
\rho+\chi\Theta^{0}_{0}&=&\frac{1}{r^{2}}-e^{-\lambda}\left(\frac{1}{r^{2}}-
\frac{\lambda'}{r}\right)-\sigma F_{1},\\\label{14}
p-\chi\Theta^{1}_{1}&=&e^{-\lambda}\left(\frac{\mu'}{r}
+\frac{1}{r^{2}}\right)-\frac{1}{r^{2}}-\sigma F_{2},\\\label{15}
p-\chi\Theta^{2}_{2}&=&e^{-\lambda}\left(\frac{\mu''}{2}
-\frac{\lambda'}{2r}+\frac{\mu'}{2r}-\frac{\mu'\lambda'}{4}
+\frac{\mu^{'2}}{4}\right)-\sigma F_{3},
\end{eqnarray}
where $F_{1},~F_{2}$ and $F_{3}$ contain the modified terms as
\begin{eqnarray}\nonumber
F_{1}&=&-\frac{R^2}{2}+2e^{-\lambda}\left[R''-\left(\frac{\lambda'}{2}
-\frac{2}{r}\right)R'\right]-2R\left[e^{-\lambda}\left(\frac{\lambda'}{r}
-\frac{1}{r^2}\right)+\frac{1}{r^2}\right],\\\nonumber
F_{2}&=&\frac{R^2}{2}-2e^{-\lambda}\left(\frac{\mu'}{2}+\frac{2}{r}\right)
R'-2R\left[e^{-\lambda}\left(\frac{\mu'}{r}+\frac{1}{r^2}\right)-\frac{1}{r^2}\right],\\\nonumber
F_{3}&=&\frac{R^2}{2}+2e^{-\lambda}\left[R'\left(\frac{\lambda'-\mu'}{2}
-\frac{1}{r}\right)-R\left(\frac{\mu''}{2}-\frac{\lambda'-\mu'}{2r}+\frac{\mu
'^2-\lambda ' \mu '}{4}\right)\right.\\\nonumber
&-&\left.R''\right].
\end{eqnarray}
We identify the effective energy density and effective pressure
components as
\begin{equation}\label{16}
{\widetilde{\rho}}^{~eff}=\rho+\chi\Theta^{0}_{0},\quad
{\widetilde{p}}^{~eff}_{r}=p-\chi\Theta^{1}_{1},\quad
{\widetilde{p}}^{~eff}_{t}=p-\chi\Theta^{2}_{2}.
\end{equation}
It is clear through direct analysis that the addition of new source
generates anisotropy in self-gravitating systems. The effective
anisotropic parameter ${\widetilde{\Delta}}^{eff}$ in the interior
of stellar objects is defined as
\begin{equation}\label{17}
{\widetilde{\Delta}}^{eff}={\widetilde{p}}^{~eff}_{t}
-{\widetilde{p}}^{~eff}_{r}=\chi(\Theta^{1}_{1}-\Theta^{2}_{2}),
\end{equation}
which vanishes if we set $\chi=0$. The system of three differential
equations (\ref{13})-(\ref{15}) interlink seven unknowns $(\mu,
\lambda, \rho, p, \Theta^{0}_{0}, \Theta^{1}_{1}, \Theta^{2}_{2})$.
In order to compute these unknowns, we follow a systematic scheme
proposed by Ovalle \cite{24}.

\section{The Extended Geometric Deformation Approach}

In this section, we formulate a solution of non-linear field
equations through the EGD approach \cite{24}. This scheme transforms
the system of field equations corresponding to the additional source
$\Theta^{\xi}_{\eta}$ into a system of quasi-field equations. The
effects of additional source $\Theta^{\xi}_{\eta}$ are analyzed by
applying geometric deformation on the metric functions ($\mu$ and
$\lambda$) as
\begin{equation}\label{18}
e^{-\lambda}=\nu+\chi h(r),\quad \mu=\alpha+\chi g(r),
\end{equation}
where the temporal and radial deformation functions are represented
by $g(r)$ and $h(r)$, respectively. Plugging these decompositions in
Eqs.(\ref{13})-(\ref{15}), we split them into two arrays. The first
system is obtained for $\chi=0$ and provides the following standard
field equations
\begin{eqnarray}\label{19}
\rho&=&\frac{1}{r^{2}}-\frac{\nu}{r^{2}}-\frac{\nu'}{r}-\frac{1}{8r^{4}}\sigma
Y_{1},\\\label{20} p&=&{\nu}\left(\frac{\alpha'}{r}
+\frac{1}{r^{2}}\right)-\frac{1}{r^{2}}+\frac{1}{8r^{4}}\sigma
Y_{2},\\\label{21} p&=&{\nu}\left(\frac{\alpha''}{2}
+\frac{\alpha'}{2r}+\frac{\alpha^{'2}}{4}\right)+\frac{\nu'}{2r}
+\frac{\nu'\alpha'}{4}+\frac{1}{8r^{4}}\sigma Y_{3},
\end{eqnarray}
where $Y_{1}$, $Y_{2}$ and $Y_{3}$ contain the modified terms
defined in Appendix A. The second set of equations, comprising of
the additional source, leads to
\begin{eqnarray}\label{23}
\Theta^{0}_{0}&=&-\frac{h}{r^{2}}-\frac{h'}{r}+\frac{1}{8r^{4}}\sigma
Z_{1},\\\label{24} \Theta^{1}_{1}&=&-h\left(\frac{\alpha'+\chi
g'}{r}+\frac{1} {r^2}\right)-\frac{\nu g'}{r}+\frac{1}{8r^{4}}\sigma
Z_{2},\\\nonumber \Theta^{2}_{2}&=&-h\left(\frac{\alpha''+\chi
g''}{2}+\frac{{(\alpha '+\chi g')}^{2}}{4} +\frac{\alpha '+\chi
g'}{2 r}\right)-\frac{h'}{2r}-\nu
\left(\frac{g''}{2}\right.\\\label{25}
&+&\left.\frac{\alpha'g'}{2}+\frac{\chi g'^2}{4}+\frac{g'}{2
r}\right)-\frac{h'}{4}  \left(\alpha '+\chi g'\right)-\frac{\nu
'g'}{4}-\frac{1}{8r^{4}}\sigma Z_{3}.
\end{eqnarray}
The terms $Z_{1}$, $Z_{2}$ and $Z_{3}$, appearing due to the
function $f(R)$, are expressed in Appendix A.

The Bianchi identity is preserved for the perfect fluid distribution
in the $(\alpha,\nu)$-frame as
\begin{equation}\label{28}
\nabla^{(\alpha,\nu)}_{\xi}{\widetilde{T}}^{\xi}_{\eta}=0,
\end{equation}
while the divergence of ${\widetilde{T}}^{\xi}_{\eta}$ associated
with metric (\ref{7}) turns out to be
\begin{equation}\label{29}
\nabla_{\xi}{\widetilde{T}}^{\xi}_{\eta}=\nabla^{(\alpha,\nu)}_{\xi}
{\widetilde{T}}^{\xi}_{\eta}+\frac{\chi
g'}{2}(\rho+p)\delta^{1}_{\eta}.
\end{equation}
For the gravitational source $\Theta^{\xi}_{\eta}$, the conservation
equation takes the form
\begin{equation}\label{30}
\nabla_{\xi}\Theta^{\xi}_{\eta}=-\frac{g'}{2}(\rho+p)\delta^{1}_{\eta}.
\end{equation}
We conclude from Eqs.(\ref{29}) and (\ref{30}) that the matter
sources (perfect fluid source and the additional source) exchange
energy in contrast to the MGD scheme where interaction is purely
gravitational. Here, it is noteworthy that the EGD technique is
applicable when there is no exchange of energy in two particular
scenarios: vacuum ($\widetilde{T}^{\xi}_{\eta}=0$) and barotropic
($\widetilde{T}^{0}_{0}=\widetilde{T}^{1}_{1}$) fluid distributions.

\section{Junction Conditions}

In order to investigate the physical features of self-gravitating
system, the junction conditions must be fulfilled at the
hypersurface $(\Sigma)$ of stellar body. A hypersurface is a
boundary between interior and exterior spacetimes that separates
them from each other. These matching conditions describe a
connection between interior and exterior spacetimes at
$r=\mathcal{R}$, where $\mathcal{R}$ denotes the radius of stellar
body. In GR, the exterior vacuum of a static spherical object is
represented by the Schwarzschild spacetime. However, in the present
work, the contributions from $f(R)$ gravity as well as the deformed
metric potentials may modify the exterior manifold. The vacuum
solution in $f(R)$ theory coincides with the Schwarzschild metric if
the function $f(R)$ belongs to class $C^{3}$ (a function whose first
three derivatives are continuous) with \cite{45*}
\begin{equation}\label{4f}
f(0)=0, \quad f_{R}(0)\neq0.
\end{equation}
The model $f(R)=R+\sigma R^2$ is consistent with these conditions.
Thus, we can use Schwarzschild spacetime to represent the exterior
vacuum.

We consider the interior geometry as
\begin{equation}\label{4a}
ds^2_{-}=e^{\mu_{-}(r)}dt^2-\left(1-\frac{2m(r)}{r}+\chi
h(r)\right)^{-1}dr^2-r^2(d\theta^2+\sin^2\theta d\phi^2),
\end{equation}
where $m(r)$ is the mass of interior geometry. The line element
describing the exterior geometry takes the following form
\begin{eqnarray}\nonumber
ds^2_{+}&=&\left(1-\frac{2\bar{m}(r)}{r}+\chi
g^{*}(r)\right)dt^2-\left(1-\frac{2\bar{m}(r)}{r}+\chi
h^{*}(r)\right)^{-1}dr^2-r^2\\\label{4e}
&\times&\left(d\theta^2+\sin^2\theta d\phi^2\right).
\end{eqnarray}
Here, subscript ``$+$'' represents the exterior spacetime. Moreover,
$\bar{m}$, $h^{*}$ and $g^{*}$ represent the exterior mass, radial
and temporal geometric deformations in the exterior Schwarzschild,
respectively. The smooth matching between interior and exterior
spacetimes at the hypersurface $\Sigma:r=\mathcal{R}$ specifies the
unknown constants. The continuity of the first fundamental form
($[ds^{2}]_{\Sigma}=0$) yields
\begin{eqnarray}\label{4c}
\alpha(\mathcal{R})+\chi
g(\mathcal{R})=1-\frac{2\bar{m}(\mathcal{R})}{\mathcal{R}}+\chi
g^{*}(\mathcal{R}),\\\label{4ca}
1-\frac{2m(\mathcal{R})}{\mathcal{R}}+\chi
h(\mathcal{R})=1-\frac{2\bar{m}(\mathcal{R})}{\mathcal{R}}+\chi
h^{*}(\mathcal{R}).
\end{eqnarray}
The second fundamental form of continuity, expressed as
$[T_{\xi\eta}^{tot}X^{\eta}]_{\Sigma}=0$ ($X^{\eta}$ is a unit four
vector in radial direction), leads to
\begin{equation}\nonumber
p(\mathcal{R})-\chi\Theta^{1}_{1}(\mathcal{R})_{-}=-\chi\Theta^{1}_{1}(\mathcal{R})_{+}.
\end{equation}
Using Eq.(\ref{24}), the above expression is rewritten as
\begin{eqnarray}\nonumber
p(\mathcal{R})+\chi\left(
h(\mathcal{R})\left(\frac{1}{\mathcal{R}^{2}}+\frac{\mu'}{\mathcal{R}}\right)+\frac{\nu
g'}{\mathcal{R}}-\frac{\sigma}{8r^{4}} Z_{2}\right)&=&\frac{\chi
h^{*}(\mathcal{R})}{\mathcal{R}^{2}}\left(1+\frac{2\bar{m}}{\mathcal{R}-2\bar{m}}\right)\\\label{4d}
&+&\frac{\chi\nu{g^{*}}'}{\mathcal{R}}-\frac{\chi\sigma}{8r^{4}}
Z_{2}^{*},
\end{eqnarray}

We assume that $h^{*}=g^{*}=0$ so that the exterior manifold reduces
to Schwarzschild metric and the pressure remains unaffected at the
boundary of the star, i.e.,
\begin{equation}\label{4g}
\widetilde{p}^{~eff}(\mathcal{R})=p(\mathcal{R})+\chi\left(
h(\mathcal{R})\left(\frac{1}{\mathcal{R}^{2}}+\frac{\mu'}{\mathcal{R}}\right)+\frac{\nu
g'}{\mathcal{R}}-\frac{\sigma}{8r^{4}} Z_{2}\right)=0.
\end{equation}
In $f(R)$ gravity, two additional conditions related to Ricci scalar
must hold to ensure a smooth junction between interior and exterior
manifolds \cite{60}. These conditions read
\begin{equation}\label{4h}
[R]_{\Sigma}=0, \quad [\nabla_{\xi}R]_{\Sigma}=0,
\end{equation}
where scalar curvature $R$ is a function of $r$ only. The conditions
in Eq.(\ref{4h}) hold for the stellar model constructed in the
current work.

\section{Anisotropic Interior Solutions}

In order to solve the system of field equations associated with the
anisotropic distribution, we need known isotropic solutions. Thus,
we choose Krori-Barua solution for perfect matter configuration
\cite{50}. This solution is known for its singularity free nature
and was initially used to study charged relativistic objects.
However, later on, this solution has also been used in the absence
of charge in GR as well as in other modified theories \cite{44*}.
The Krori-Barua solution is isotropic in the presence of
electromagnetic field but it may not correspond to an isotropic
spacetime in the absence of charge. The metric potentials of
Krori-Barua solution generate a purely isotropic fluid distribution
in $f(R)$ theory, if and only if $p_{r}=p_{t}$. Therefore, in order
to evaluate the expression of isotropic pressure, we employ this
condition $p_{r}=p_{t}$. Thus, the Krori-Barua solution in $f(R)$
gravity takes the form
\begin{eqnarray}\label{34}
e^{\alpha(r)}&=&e^{\mathcal{B}r^{2}+\mathcal{C}},\\\label{35}
e^{\lambda(r)}&=&\nu^{-1}(r)=e^{\mathcal{A}r^{2}},\\\nonumber
\rho&=&e^{-\mathcal{A} r^2} \left(2
\mathcal{A}-\frac{1}{r^2}\right)+\frac{1}{r^2}-\frac{2 \sigma e^{-2
\mathcal{A} r^2}}{r^4} \left[e^{2 \mathcal{A} r^2}-\mathcal{B}^4
r^8-6 \mathcal{B}^3 r^6\right.\\\nonumber
&-&\left.6e^{\mathcal{A}r^2}+3 \mathcal{B}^2 r^4+5-12 \mathcal{A}^3
r^6 \left(\mathcal{B} r^2+2\right) +\mathcal{A}^2 r^4 \left(11
\mathcal{B}^2 r^4\right.\right.\\\label{36}
&+&\left.\left.68\mathcal{B}r^2+40\right)+2 \mathcal{A} r^2
\left(\mathcal{B}^3 r^6-13 \mathcal{B}^2 r^4-24 \mathcal{B}
r^2+2\right)\right],\\\nonumber
p&=&\frac{e^{-2\mathcal{A}r^2}}{2r^2}\left[-e^{2\mathcal{A}r^2}
+e^{\mathcal{A}r^2}\left\{1+\mathcal{B}^2r^4-\mathcal{A}\left(\mathcal{B}r^4
+r^2+12\sigma\right)+4\mathcal{B}\left(r^2\right.\right.\right.\\\nonumber
&+&\left.\left.\left.3\sigma\right)\right\}-4\sigma\left\{6
\mathcal{A}^3r^4\left(\mathcal{B}r^2+2\right)+\mathcal{A}\left(4\mathcal{B}^3
r^6+34\mathcal{B}^2r^4+35\mathcal{B}r^2-3\right)\right.\right.\\\label{37}
&-&\left.\left.\mathcal{A}^2 r^2 \left(10 \mathcal{B}^2 r^4+43
\mathcal{B} r^2+20\right)+\mathcal{B} \left(-5 \mathcal{B}^2 r^4-11
\mathcal{B} r^2+3\right)\right\}\right],
\end{eqnarray}
where $\mathcal{A}$, $\mathcal{B}$ and $\mathcal{C}$ are constants
that can be computed through matching conditions on the
hypersurface. The smooth matching of external and internal regions
on the hypersurface determine the unknown constants of the
anisotropic solution and contribute to the investigation of its
physical features. Here, we consider Schwarzschild as an exterior
spacetime described by the line element
\begin{equation}\label{38}
ds^2_{+}=(1-\frac{2\bar{m}(r)}{r})dt^2-\frac{1}{(1-\frac{2\bar{m}(r)}{r})}dr^2
-r^2(d\theta^2+\sin^2\theta d\phi^2),
\end{equation}
The continuity of the metric components $g_{00},~g_{11}$ and
$g_{00,1}$ at the boundary ($r=\mathcal{R}$ and total
mass$=\mathcal{M}$) yields
\begin{eqnarray}\label{39}
\mathcal{A}&=&-\frac{1}{\mathcal{R}^{2}}\ln\left(1-\frac{2\mathcal{M}}
{\mathcal{R}}\right),\\\label{40}
\mathcal{B}&=&\frac{\mathcal{M}}{\mathcal{R}^{2}(\mathcal{R}
-2\mathcal{M})},\\\label{41}
\mathcal{C}&=&\ln\left(1-\frac{2\mathcal{M}}{\mathcal{R}}\right)-\frac
{\mathcal{M}}{\mathcal{R}-2\mathcal{M}}.
\end{eqnarray}
with the compactness parameter
$\frac{\mathcal{M}}{\mathcal{R}}<\frac{4}{9}\left(1+\frac{\beta}{6}\right)$,
where $\beta$ (with $0\leq\beta\ll1$) denotes small modification in
the Buchdahl-Bondi limit \cite{45}. The model is also consistent
with the conditions in Eq.(\ref{4h}). For anisotropic model, the
expressions of effective matter variables are evaluated as follows
\begin{eqnarray}\label{42*}
{\widetilde{\rho}}^{~eff}&=&\frac{1}{r^{2}}-\frac{\nu}{r^{2}}-\frac{\nu'}{r}-\frac{1}{8r^{4}}\sigma
Y_{1}+\chi\left[-\frac{h}{r^{2}}-\frac{h'}{r}+\frac{1}{8r^{4}}\sigma
Z_{1}\right],\\\nonumber
{\widetilde{p}}_{r}^{~eff}&=&{\nu}\left(\frac{\alpha'}{r}
+\frac{1}{r^{2}}\right)-\frac{1}{r^{2}}+\frac{1}{8r^{4}}\sigma
Y_{2}-\chi\left[-h\left(\frac{\alpha'+\chi g'}{r}+\frac{1}
{r^2}\right)-\frac{\nu g'}{r}\right.\\\label{42**}
&+&\left.\frac{1}{8r^{4}}\sigma Z_{2}\right],\\\nonumber
{\widetilde{p}}_{t}^{~eff}&=&{\nu}\left(\frac{\alpha''}{2}
+\frac{\alpha'}{2r}+\frac{\alpha^{'2}}{4}\right)+\frac{\nu'}{2r}
+\frac{\nu'\alpha'}{4}+\frac{1}{8r^{4}}\sigma
Y_{3}-\chi\left[-h\left(\frac{\alpha''+\chi
g''}{2}\right.\right.\\\nonumber &+&\left.\left.\frac{{(\alpha'+\chi
g')}^{2}}{4} +\frac{\alpha'+\chi g'}{2 r}\right)-\frac{h'}{2r}-\nu
\left(\frac{g''}{2} +\frac{\alpha'g'}{2}+\frac{\chi
g'^2}{4}+\frac{g'}{2 r}\right)\right.\\\label{42***}
&-&\left.\frac{h'}{4}\left(\alpha'+\chi g'\right)-\frac{\nu
'g'}{4}-\frac{1}{8r^{4}}\sigma Z_{3}\right],
\end{eqnarray}
with anisotropic factor
\begin{eqnarray}\nonumber
\widetilde{\Delta}^{eff}&=&\chi\left[-h\left(\frac{\alpha'+\chi
g'}{r}+\frac{1} {r^2}\right)-\frac{\nu g'}{r}+\frac{1}{8r^{4}}\sigma
Z_{2}-\left\{-h\left(\frac{\alpha''+\chi
g''}{2}\right.\right.\right.\\\nonumber
&+&\left.\left.\left.\frac{{(\alpha'+\chi g')}^{2}}{4}
+\frac{\alpha'+\chi g'}{2 r}\right)-\frac{h'}{2r}-\nu
\left(\frac{g''}{2} +\frac{\alpha'g'}{2}+\frac{\chi
g'^2}{4}+\frac{g'}{2 r}\right)\right.\right.\\\label{42****}
&-&\left.\left.\frac{h'}{4}\left(\alpha'+\chi g'\right)-\frac{\nu
'g'}{4}-\frac{1}{8r^{4}}\sigma Z_{3}\right\}\right].
\end{eqnarray}

The system (\ref{23})-(\ref{25}) interlinks the components of
additional source with the deformation functions. In order to solve
this system of quasi-field equations, we need additional constraints
to close the system. In this regard, we implement a barotropic
equation of state on $\Theta^{\xi}_{\eta}$ as
\begin{equation}\label{42}
\Theta^{0}_{0}=\delta\Theta^{1}_{1}+\gamma\Theta^{2}_{2},
\end{equation}
For simplicity, we set $\gamma=0$ and $\delta=1$ which forms the
relation $\Theta^{0}_{0}=\Theta^{1}_{1}$. We also employ two
additional constraints (density-like and pressure-like) and
formulate corresponding solutions.

\subsection{Solution I}

In this section, we apply an additional constraint on the temporal
component of source term in order to close the system. We adopt
density-like constraint such as
\begin{equation}\label{43}
\Theta^{0}_{0}=\rho.
\end{equation}
The deformation functions are evaluated numerically through
Eqs.(\ref{42}) and (\ref{43}) by plugging the metric potentials of
Krori-Barua spacetime. For this constraint, the presence of higher
order derivatives of unknown functions hinders the extraction of
numerical solutions. Therefore, we assume quadratic deformation
functions and solve the corresponding system of differential
equations for the initial conditions
$g(0.01)=g'(0.01)=h(0.01)=h'(0.01)=0$. The values of constants
$\mathcal{A}$ and $\mathcal{B}$ evaluated in Eq.(\ref{39}) and
(\ref{40}) are used to evaluate the numerical solution.
\begin{figure}\center
\epsfig{file=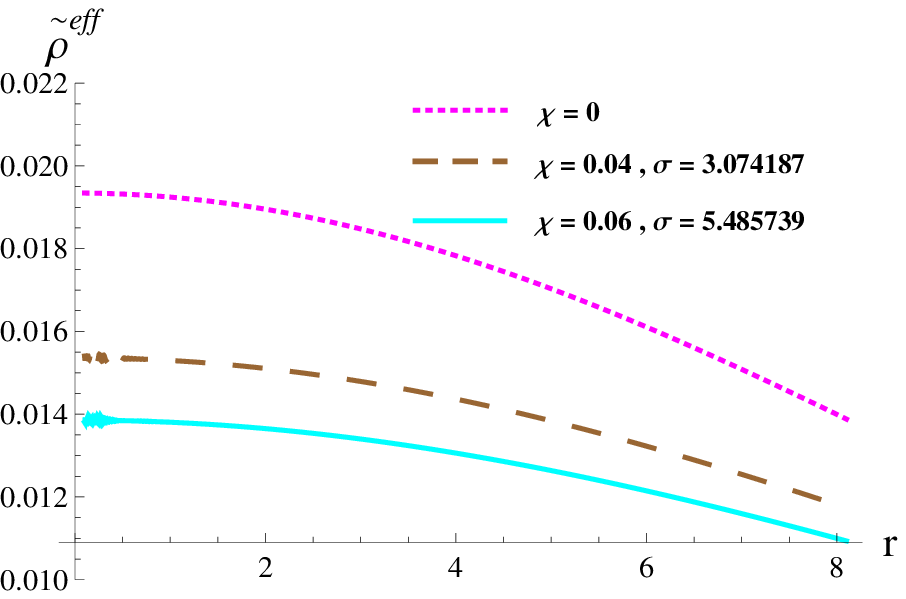,width=0.4\linewidth}\epsfig{file=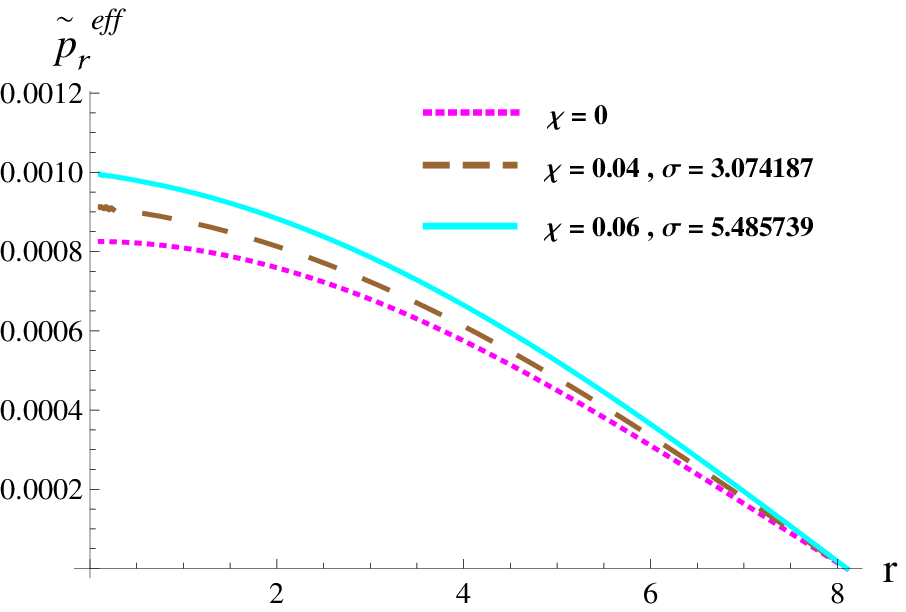,width=0.4\linewidth}
\epsfig{file=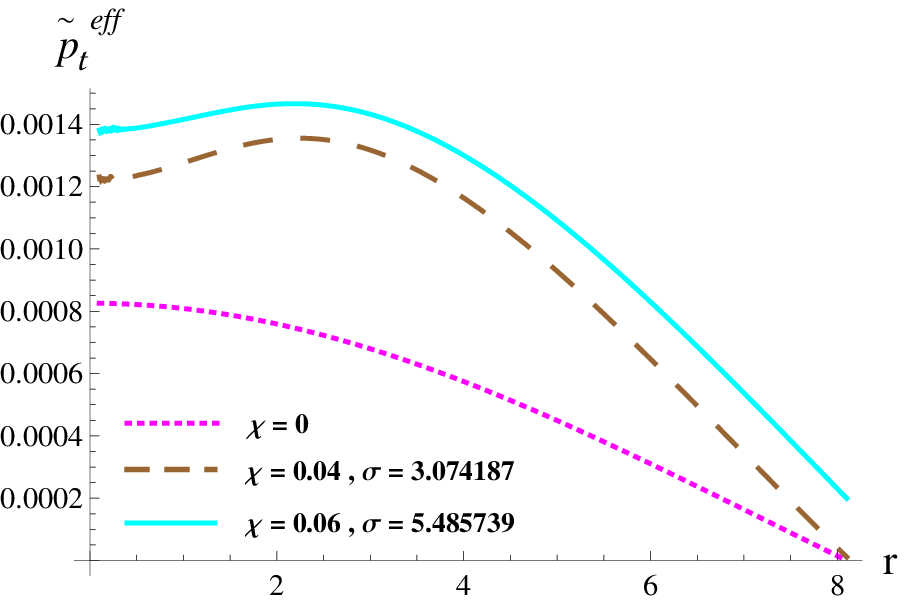,width=0.4\linewidth}\epsfig{file=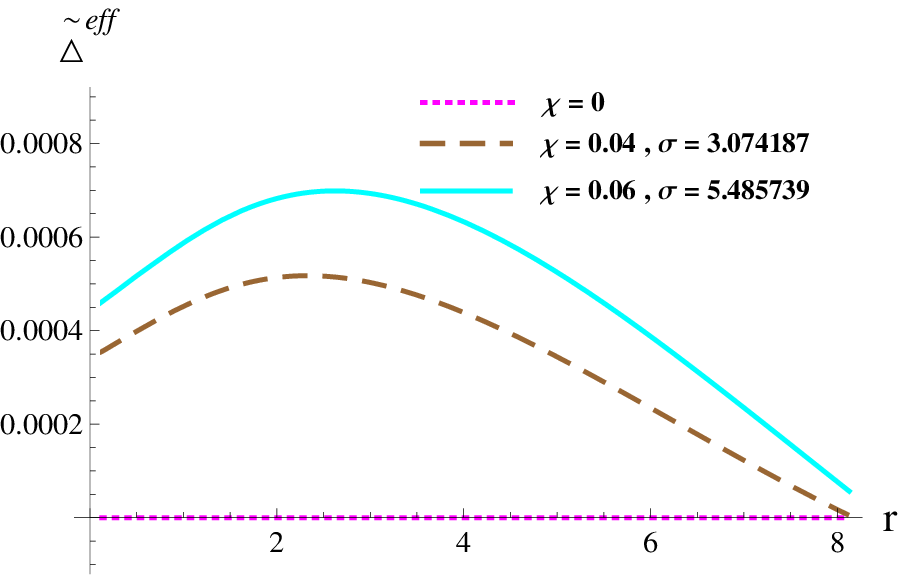,width=0.4\linewidth}\\
\caption{Plots of ${\widetilde{\rho}}^{~eff}$,
${\widetilde{p}}^{~eff}_{r}$, ${\widetilde{p}}^{~eff}_{t}$ and
${\widetilde{\Delta}}^{eff}$ for the solution I.} \label{43a}
\end{figure}
We discuss physical characteristics of the stellar bodies through
graphical analysis corresponding to the star Her X-I with radius
$\mathcal{R}=8.10km$ and mass $\mathcal{M}=1.25375km$ \cite{46}. We
compute values of $\sigma$ for $\chi=0.04,~0.06$ through the
condition $\widetilde{p}^{~eff}_r(\mathcal{R})=0$.

The energy density and pressure (radial as well as temporal) of a
well-behaved stellar structure must be finite, positive and maximum
at the center. The plots of physical parameters
(${\widetilde{\rho}}^{~eff}$, ${\widetilde{p}}^{~eff}_{r}$,
${\widetilde{p}}^{~eff}_{t}$) along with anisotropy factor are
displayed in Figure \textbf{\ref{43a}}. The profile of effective
energy density indicates that it is maximum at $r=0$ and declines
gradually with increasing $r$. It is found that an increase in
$\chi$ and $\sigma$ causes a decrease in
${\widetilde{\rho}}^{~eff}$. The graph of effective radial pressure
${\widetilde{p}}^{~eff}_{r}$ depicts that it vanishes at the star's
surface and decreases monotonically as $\sigma$ and $\chi$ decrease.
The trend of ${\widetilde{p}}^{~eff}_{r}$ gradually decreases with
respect to $r$. It is observed that the behavior of
${\widetilde{p}}^{~eff}_{t}$ decreases towards the boundary and at
the center of the star, it increases for larger values of $\chi$.
The plot of effective anisotropic factor shows that
${\widetilde{\Delta}}^{~eff}>0$ away from the center. Moreover, it
violates the regularity condition as radial and tangential pressures
are not same at the center of the celestial object. We observe that
the anisotropic factor vanishes for $\chi=0$.
\begin{figure}\center
\epsfig{file=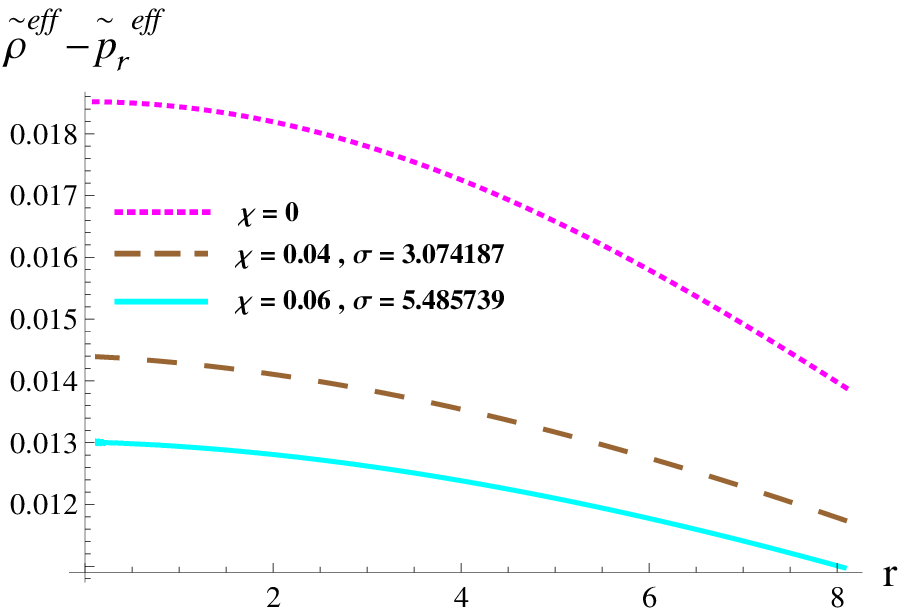,width=0.4\linewidth}\epsfig{file=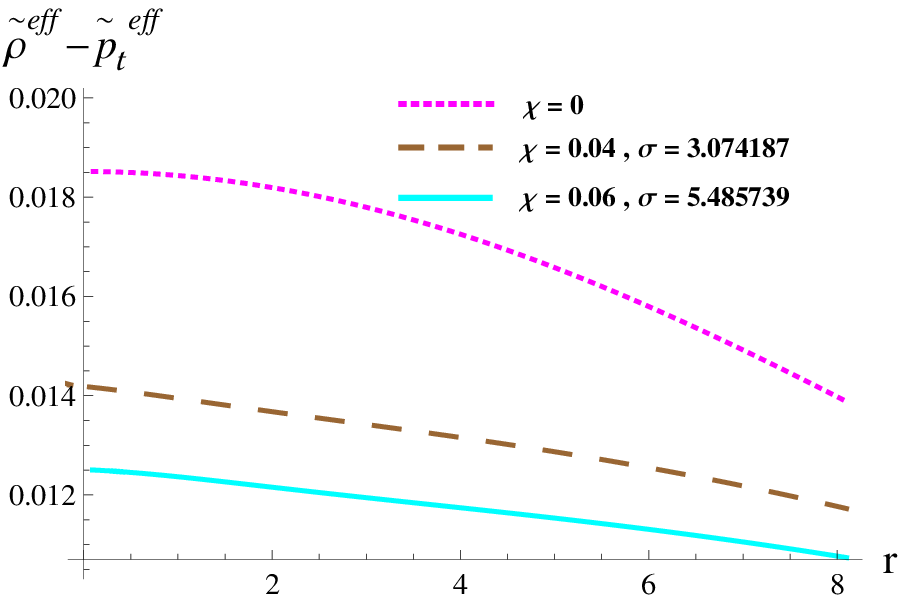,width=0.4\linewidth}\\
\caption{Plots of $\widetilde{\rho}^{~eff}-\widetilde{p}_r^{~eff}$
and $\widetilde{\rho}^{~eff}-\widetilde{p}_{t}^{~eff}$ for solution
I.} \label{43b}
\end{figure}

To measure the viability of the resulting solution, four energy
conditions (null (NEC), weak (WEC), strong (SEC) and dominant (DEC))
must be satisfied. These energy conditions indicate the presence of
ordinary matter in a compact celestial system. In $f(R)$ scenario,
these energy bounds are given in terms of $\widetilde{\rho}^{~eff}$,
$\widetilde{p}_r^{~eff}$ and $\widetilde{p}_t^{~eff}$ as
\begin{eqnarray*}
&&\text{NEC:}\quad\widetilde{\rho}^{~eff}+\widetilde{p}_r^{~eff}\geq0,
\quad\widetilde{\rho}^{~eff}+\widetilde{p}_{t}^{~eff}\geq0,\\
&&\text{WEC:}~~\widetilde{\rho}^{~eff}\geq0,\quad
\widetilde{\rho}^{~eff}+\widetilde{p}_r^{~eff}\geq0,
\quad\widetilde{\rho}^{~eff}+\widetilde{p}_{t}^{~eff}\geq0,\\
&&\text{SEC:}\quad\widetilde{\rho}^{~eff}+\widetilde{p}_r^{~eff}
+2\widetilde{p}_{t}^{~eff}\geq0,\\
&&\text{DEC:}~~\widetilde{\rho}^{~eff}-\widetilde{p}_r^{~eff}\geq0,\quad
\widetilde{\rho}^{~eff}-\widetilde{p}_{t}^{~eff}\geq0.
\end{eqnarray*}
As Figure \textbf{\ref{43a}} illustrates the positive behavior of
$\widetilde{\rho}^{~eff}$, $\widetilde{p}_r^{~eff}$ and
$\widetilde{p}_t^{~eff}$, the null, weak and strong energy
conditions are satisfied. Therefore, we only display the plots of
DEC which also exhibit positive behavior as shown in Figure
\textbf{\ref{43b}}. Hence, the graphical behavior assures the
physical viability of the constructed solution.

\begin{figure}\center
\epsfig{file=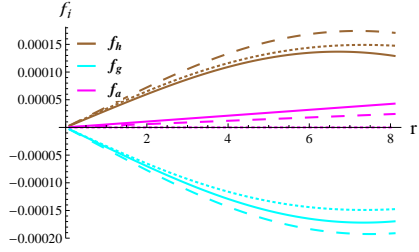,width=0.4\linewidth}\\
\caption{Plots of hydrostatic equilibrium with
$\chi=0~(\text{dotted})$, $0.04~(\text{dashed})$,
$0.06~(\text{solid})$ for solution I.}\label{43c}
\end{figure}

In order to determine the equilibrium state of the constructed
anisotropic model, we use the Tolman-Oppenheimer-Volkoff (TOV)
equation \cite{54}. This equation demonstrates that sum of all
physical forces acting on the system must be equal to zero. In the
considered setup these forces are classified as gravitational
($f_{g}$), anisotropic ($f_{a}$) and hydrostatic ($f_{h}$) forces.
Corresponding to the spherical spacetime, the TOV equation becomes
\begin{equation*}
-[p'-\chi(\Theta^{1}_{1})']-[\frac{\alpha'}{2}(\rho+p)+\frac{\chi
g'}{2}(\rho+p)+\frac{\mu'\chi}{2}(\Theta^{0}_{0}
-\Theta^{1}_{1})]+[\frac{2\chi}{r}(\Theta^{2}_{2}-\Theta^{1}_{1})]=0.
\end{equation*}
The hydrostatic, gravitational and anisotropic forces are,
respectively expressed as
\begin{eqnarray*}
f_{h}&=&-[p'-\chi(\Theta^{1}_{1})'],\\\nonumber
f_{g}&=&-[\frac{\alpha'}{2}(\rho+p)+\frac{\chi
g'}{2}(\rho+p)+\frac{\mu'\chi}{2}(\Theta^{0}_{0}
-\Theta^{1}_{1})],\\\nonumber
f_{a}&=&[\frac{2\chi}{r}(\Theta^{2}_{2}-\Theta^{1}_{1})].
\end{eqnarray*}
The graphical analysis of these forces in Figure \textbf{\ref{43c}}
exhibits that gravitational force is balanced by the remaining
forces. Moreover, for $\chi=0$, the gravitational force vanishes and
the other two force counter balance each other. This depicts that
the constructed model is in hydrostatic equilibrium. We now check
the stability of the constructed anisotropic model.
\begin{figure}\center
\epsfig{file=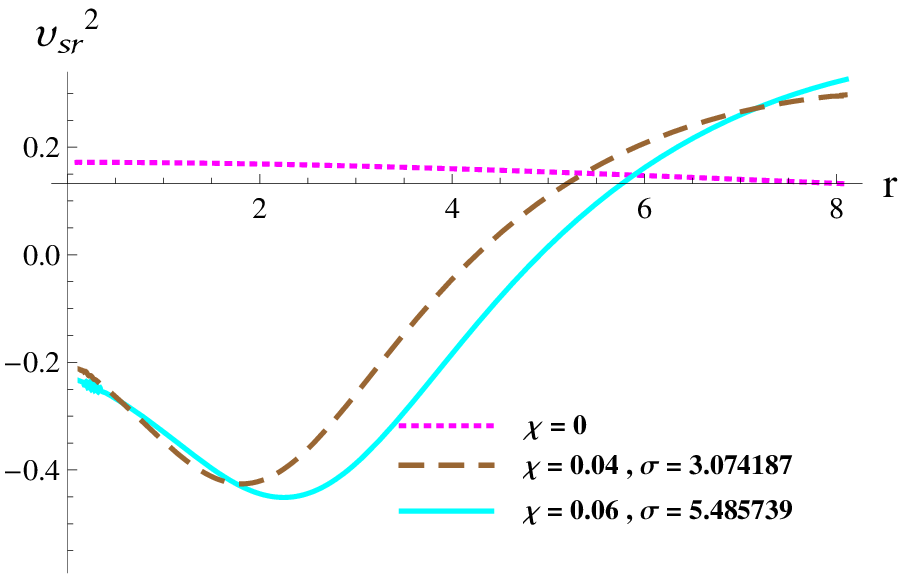,width=0.4\linewidth}\epsfig{file=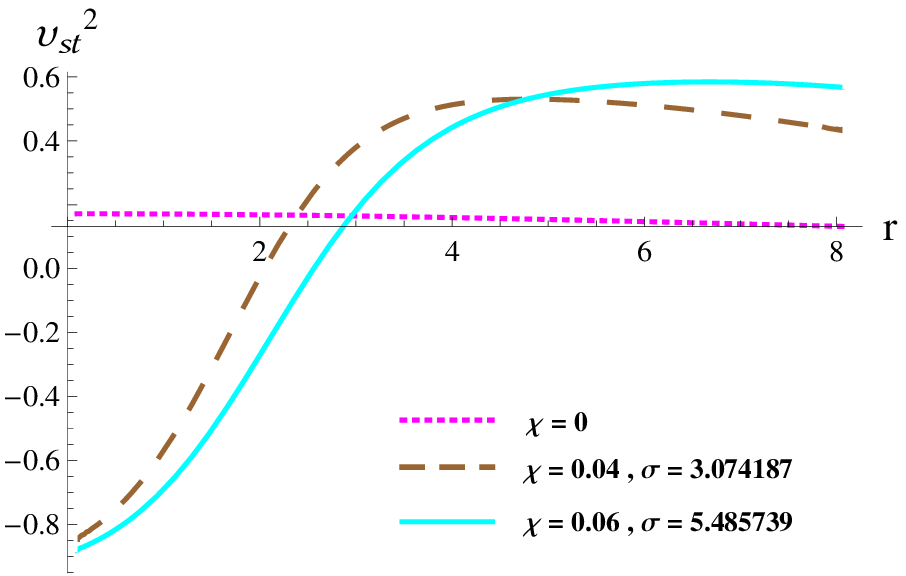,width=0.4\linewidth}
\epsfig{file=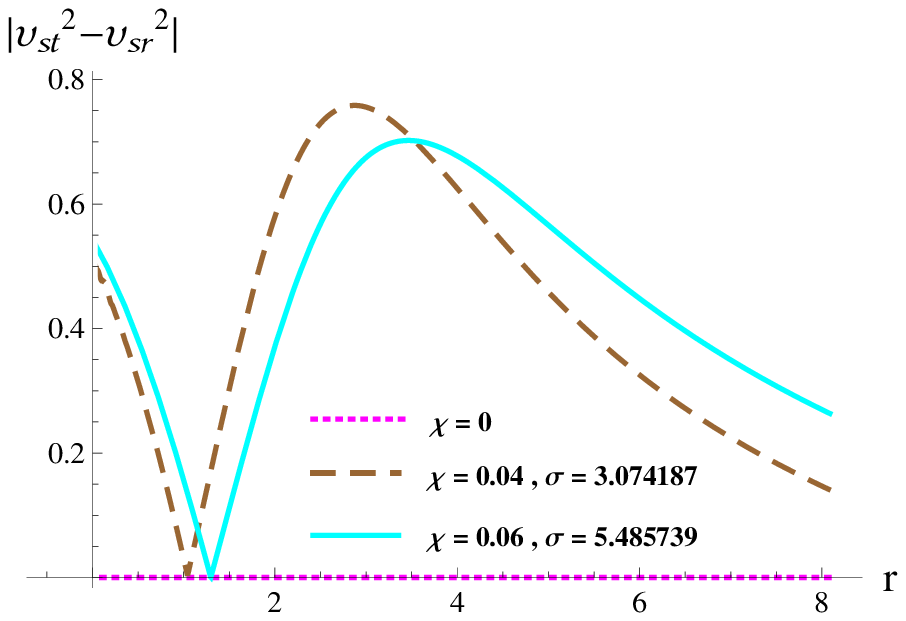,width=0.4\linewidth}\\
\caption{Plots of $v_{sr}^2$, $v_{st}^2$ and $|v_{st}^2-v_{sr}^2|$
for the solution I.}\label{43d}
\end{figure}

The stability of a self-gravitating body is an important feature
that ensures its existence and realistic matter configuration. In
this regard, causality condition \cite{48} (squared speed of sound
$v_{s}^2$ must lie in the range $[0,1]$) is a substantial tool to
examine the stability of celestial objects. Herrera \cite{49}
proposed the idea of cracking by analyzing potentially stable or
unstable regions of celestial objects. According to this idea, these
regions are defined as
\begin{eqnarray*}
&&\bullet\quad-1\leq v_{st}^2-v_{sr}^2\leq0\Longrightarrow
\text{Potentially~stable~mode}\\
&&\bullet\quad\quad0<v_{st}^2-v_{sr}^2\leq1\Longrightarrow
\text{Potentially~unstable~mode}
\end{eqnarray*}
where $v_{st}^2$ and $v_{sr}^2$ are the tangential and radial
components of squared speed of sound, respectively and are defined
as
\begin{equation*}
v_{sr}^2=\frac{d\widetilde{p}_r^{~eff}}{d\widetilde{\rho}^{~eff}},\quad
v_{st}^2=\frac{d\widetilde{p}_t^{~eff}}{d\widetilde{\rho}^{~eff}}.
\end{equation*}
The cracking conditions are expressed in combined form as
$|v_{st}^2-v_{sr}^2|<1$. Figure \textbf{\ref{43d}} illustrates the
graphical behavior of $v_{sr}^2$ and $v_{st}^2$ which indicates that
the anisotropic extension is not stable. On the other hand, cracking
condition $|v_{st}^2-v_{sr}^2|$ yields stable behavior of system for
the chosen values of parameters.

The system obeys a stiff equation of state (EoS) if an increase in
density causes an effective increase in pressure. A structure
associated with a stiff EoS is harder to compress and more stable as
compared to a setup corresponding to a soft EoS. The stiffness of
EoS is measured through adiabatic index ($\Gamma$). According to the
condition proposed by Heintzmann and Hillebrandt \cite{53}, the
adiabatic index must be greater than $\frac{4}{3}$ for a stable
model in equilibrium. However, the inclusion of local anisotropies
in the system changes the upper limit. So, in the anisotropic case,
the adiabatic index should satisfy \cite{80}
\begin{equation}\nonumber
\Gamma>\frac{4}{3}+\left[\frac{r}{3}\frac{\widetilde{\rho}_{0}^{~eff}
\widetilde{p}_{r0}^{~eff}}{|({\widetilde{p}_{r0}}^{~eff})'|}
+\frac{4}{3}\frac{(\widetilde{p}_{t0}^{~eff}-\widetilde{p}_{r0}^{~eff})}
{r|({\widetilde{p}_{r0}}^{~eff})'|}\right],
\end{equation}
where $\widetilde{p}_{r0}^{~eff}$, $\widetilde{p}_{r0}^{~eff}$ and
$\widetilde{p}_{t0}^{~eff}$ denote effective initial density,
effective initial radial and tangential pressures.  The above
expression involves the contributions from local anisotropies and
represents relativistic corrections in the adiabatic index. However,
Chandrasekhar \cite{81} pointed out that relativistic corrections to
the adiabatic index could induce instabilities within the stellar
interior. To resolve this problem, Moustakidis \cite{82} introduced
a more strict condition on $\Gamma$ and proposed a critical value of
adiabatic index ($\Gamma_{critical}$). The value of critical
adiabatic index depends on the amplitude of Lagrangian displacement
$(\zeta(r))$ from equilibrium and the compactness parameter
$2\mathcal{M}/\mathcal{R}$. Considering a particular value of the
parameter $\zeta(r)$, we obtain the critical adiabatic index as
\begin{equation*}
\Gamma_{critical}=\frac{4}{3}+\frac{19}{21}\left(\frac{\mathcal{M}}{\mathcal{R}}\right).
\end{equation*}
Thus, the stability condition becomes $\Gamma\geq\Gamma_{critical}$,
where the $\Gamma$ is defined as
\begin{equation*}
\Gamma=\frac{\widetilde{p}_r^{~eff}+\widetilde{\rho}^{~eff}}{\widetilde{p}_r^{~eff}}
\frac{d\widetilde{p}_r^{~eff}}{d\widetilde{\rho}^{~eff}}
=\frac{\widetilde{p}_r^{~eff}+\widetilde{\rho}^{~eff}}{\widetilde{p}_r^{~eff}}v_{sr}^2.
\end{equation*}
\begin{figure}\center
\epsfig{file=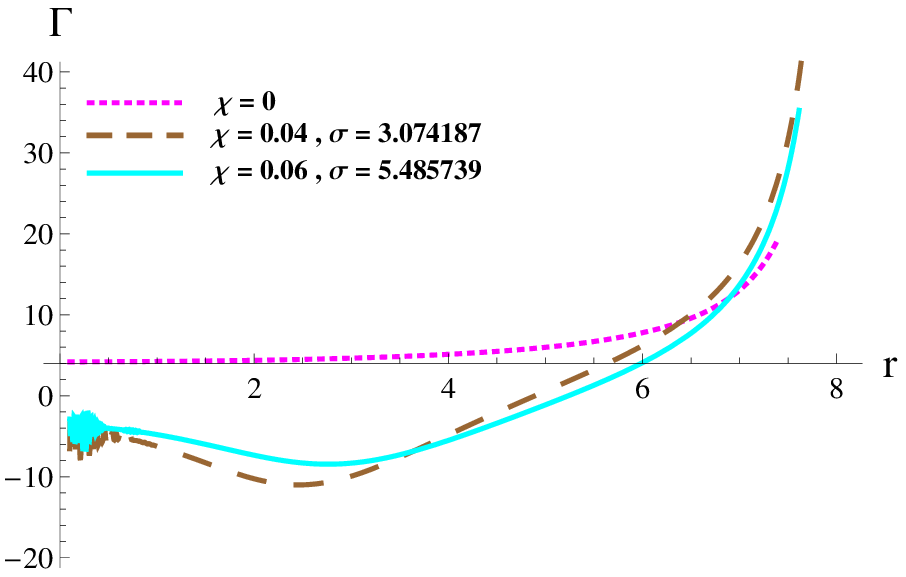,width=0.4\linewidth}\epsfig{file=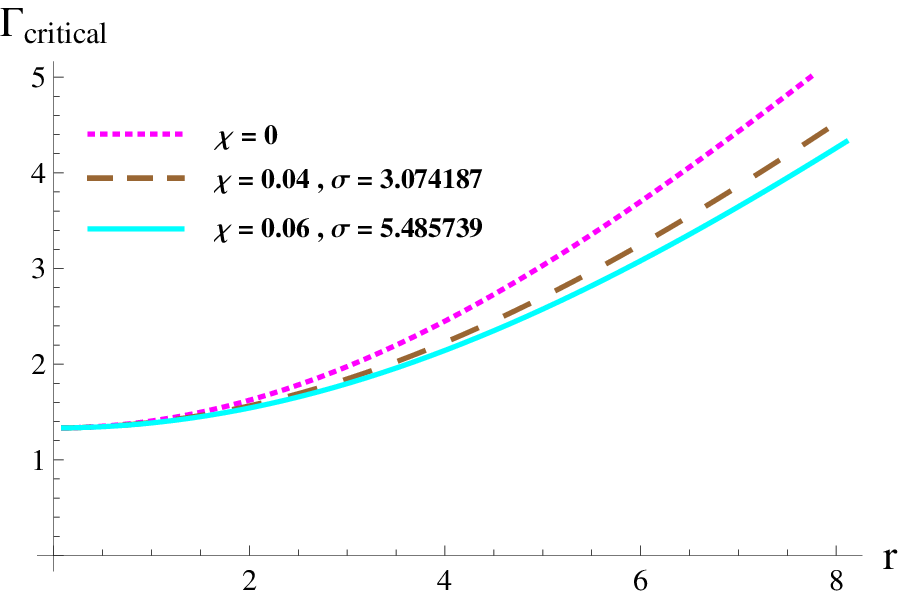,width=0.4\linewidth}\\
\caption{Plot of $\Gamma$ and $\Gamma_{critical}$ for solution
I.}\label{43e}
\end{figure}

A compact star with an increasing and positive anisotropy factor,
behaves stable for the limit given above. The positive anisotropy
generates a repulsive force that counteracts against the inward
gravitational pull.  This implies that a star does not collapse for
$\widetilde{p}_t^{~eff}>\widetilde{p}_r^{~eff}$. Thus, the analysis
of adiabatic index in radial direction is sufficient to gauge the
stability of the spherical system. The rapid decrease in the radial
pressure near the boundary of the star causes $\Gamma$ to increase
at a faster rate. Moreover, radial adiabatic index behaves
asymptotically near the star's surface as
$\widetilde{p}_r^{~eff}(\mathcal{R})=0$. The profiles of radial and
critical adiabatic index in Figure \textbf{\ref{43e}} are not
consistent with the inequality $\Gamma\geq\Gamma_{critical}$. Thus,
the extended solution is locally unstable in the presence of higher
curvature terms of $f(R)$.

\subsection{Solution II}
\begin{figure}\center
\epsfig{file=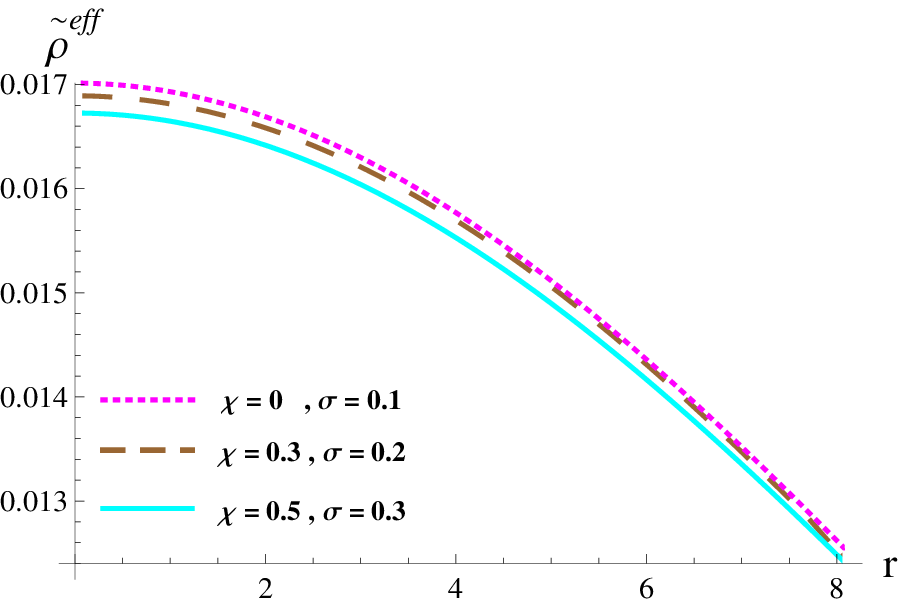,width=0.4\linewidth}\epsfig{file=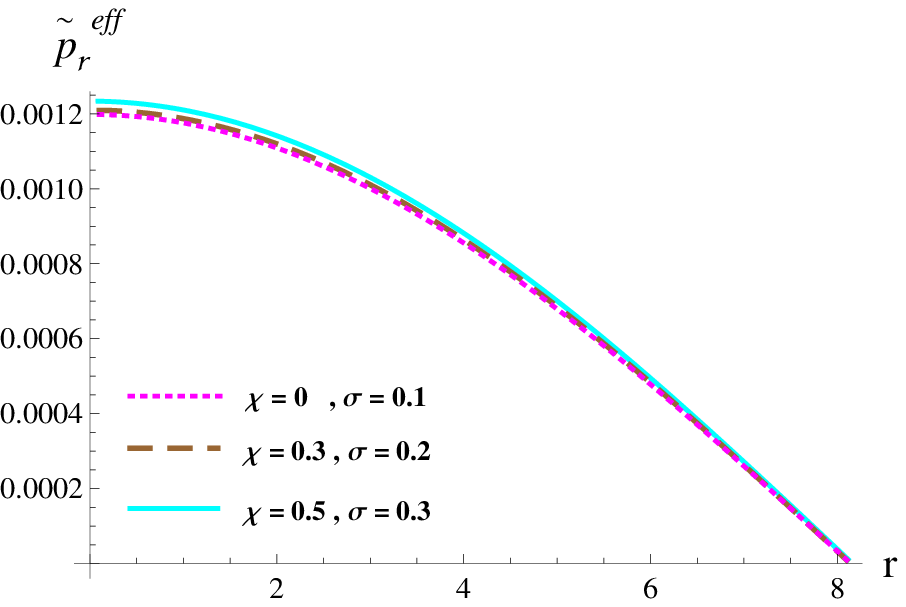,width=0.4\linewidth}
\epsfig{file=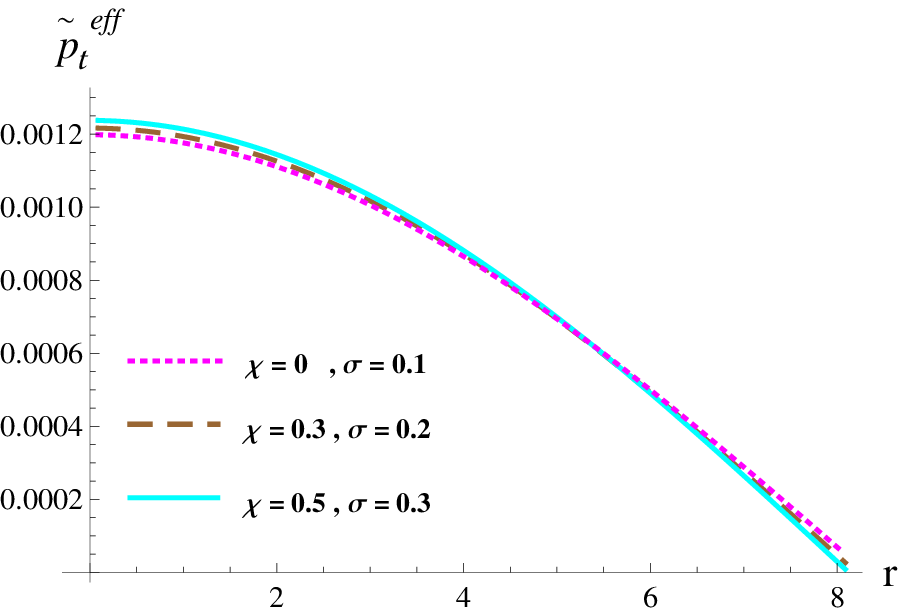,width=0.4\linewidth}\epsfig{file=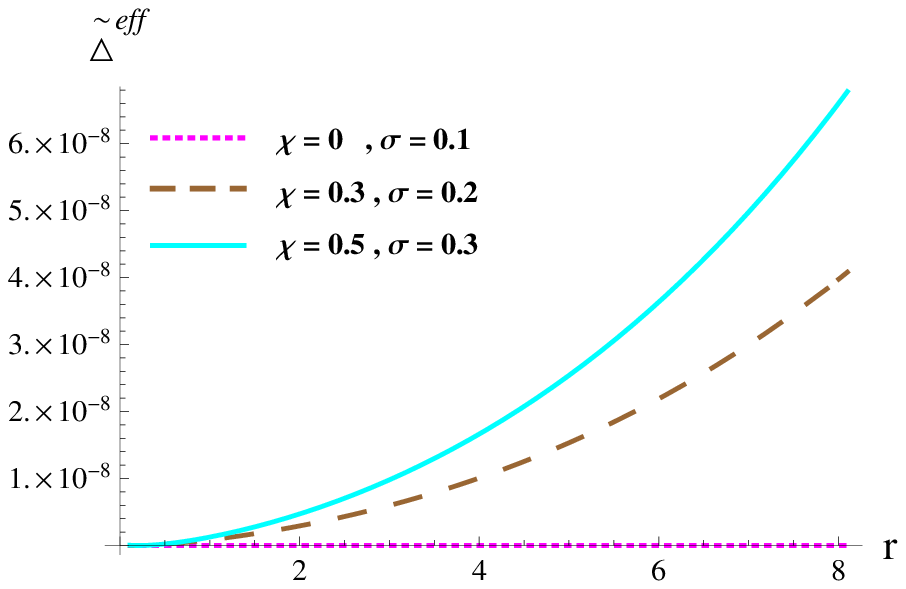,width=0.4\linewidth}\\
\caption{Plots of ${\widetilde{\rho}}^{~eff}$,
${\widetilde{p}}^{~eff}_{r}$, ${\widetilde{p}}^{~eff}_{t}$ and
${\widetilde{\Delta}}^{eff}$ for solution II.}\label{44a}
\end{figure}

In order to obtain the second anisotropic solution, we apply a
constraint on radial component of $\Theta^{\xi}_{\eta}$. The
matching of Schwarzschild exterior and deformed interior metric on
the boundary stipulates
$p(\mathcal{R})\sim\chi({\Theta^{1}_{1}}(\mathcal{R}))_{-}$. Thus,
\begin{equation}\label{54}
\Theta^{1}_{1}=p,
\end{equation}
is considered as a suitable constraint. We solve Eqs.(\ref{42}) and
(\ref{54}) simultaneously by employing Eqs.(\ref{34}), (\ref{35}),
(\ref{39}) and (\ref{40}). The unknown functions $h(r)$ and $g(r)$
are determined numerically for the initial conditions
$g(0.01)=1\times10^{-9}$, $g'(0.01)=g''(0.01)=1\times10^{-7}$,
$g'''(0.01)=1\times10^{-10}$, $h(0.01)=9\times10^{-8}$,
$h(0.01)=1\times10^{-8}$ and $h''(0.01)=1\times10^{-6}$. The
physical behavior of solution II is investigated for the same
compact star Her X-I.

\begin{figure}\center
\epsfig{file=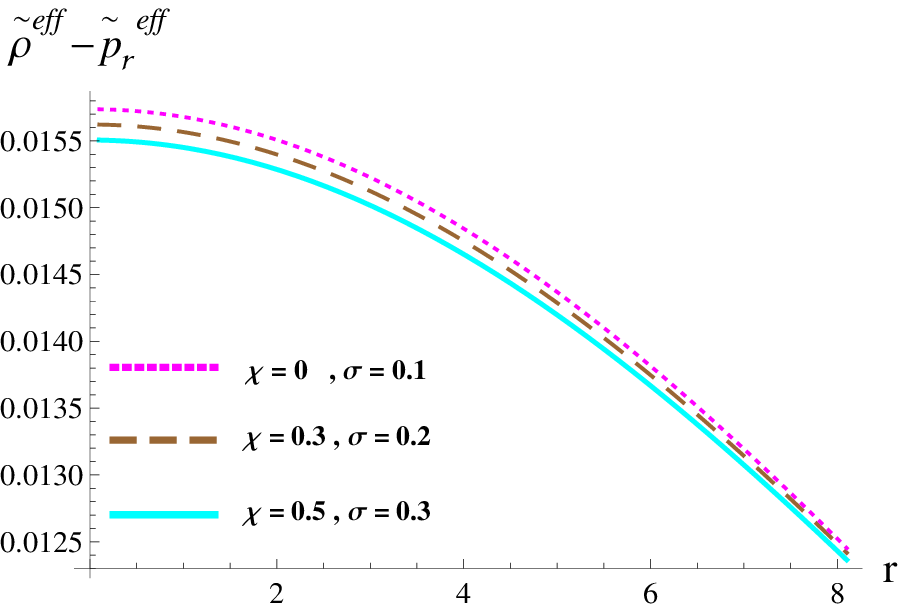,width=0.4\linewidth}\epsfig{file=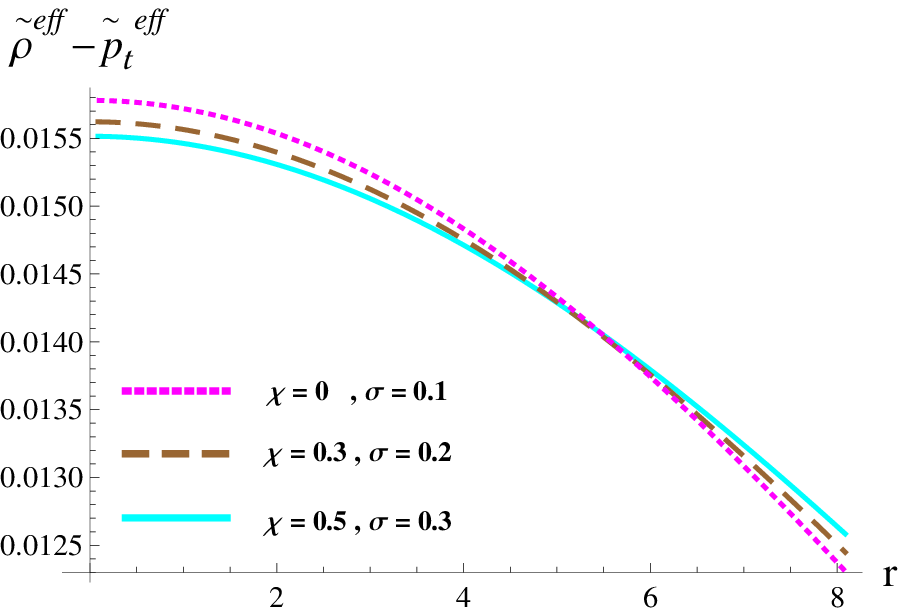,width=0.4\linewidth}\\
\caption{Plots of $\widetilde{\rho}^{~eff}-\widetilde{p}_r^{~eff}$
and $\widetilde{\rho}^{~eff}-\widetilde{p}_{t}^{~eff}$ for solution
II.}\label{44b}
\end{figure}
\begin{figure}\center
\epsfig{file=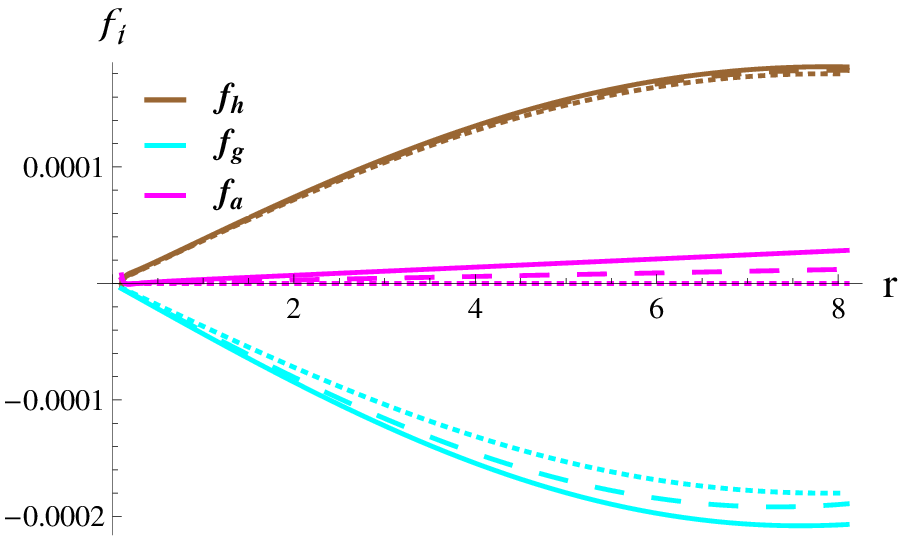,width=0.4\linewidth}\\
\caption{Plots of hydrostatic equilibrium with
$\chi=0~(\text{dotted})$, $0.3~(\text{dashed})$,
$0.5~(\text{solid})$ for solution II.}\label{44c}
\end{figure}
\begin{figure}\center
\epsfig{file=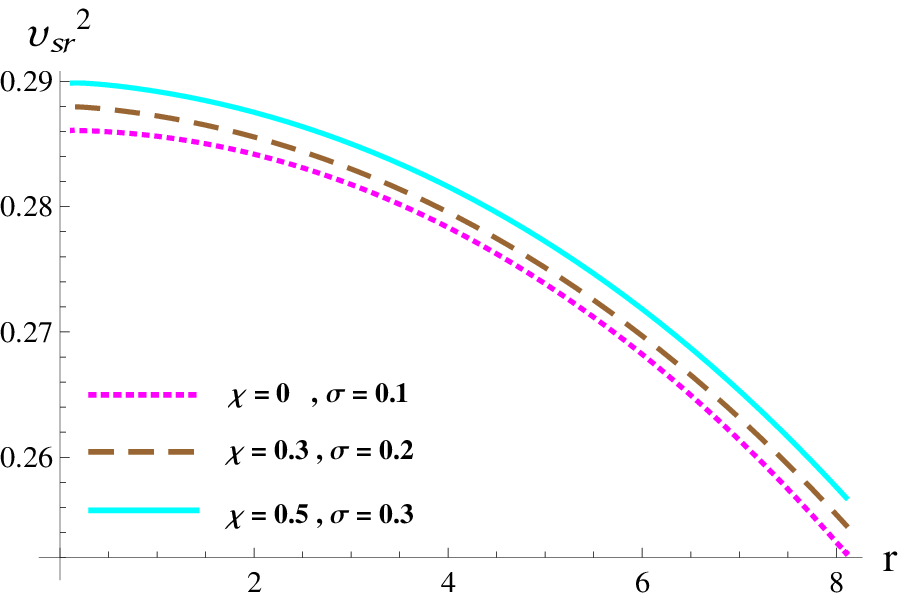,width=0.4\linewidth}\epsfig{file=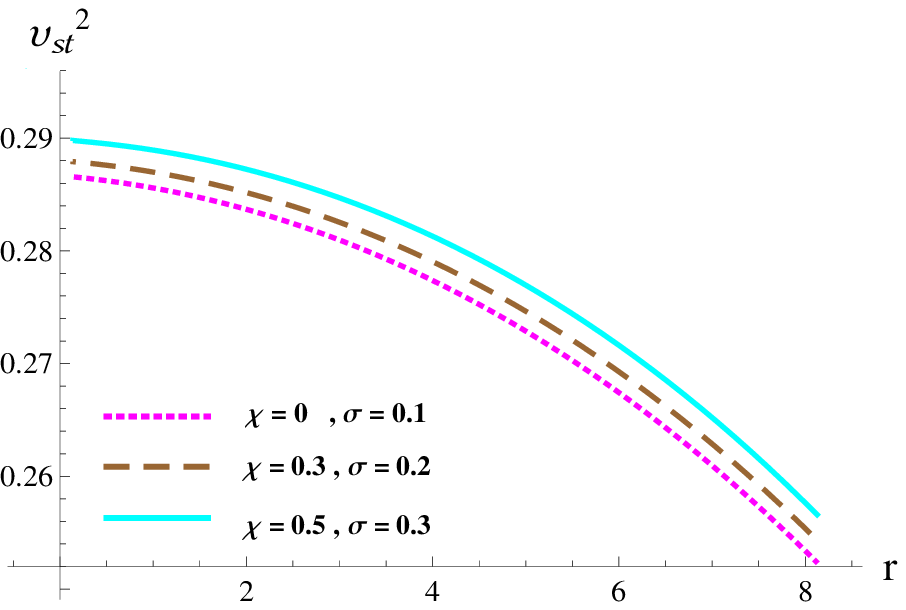,width=0.4\linewidth}
\epsfig{file=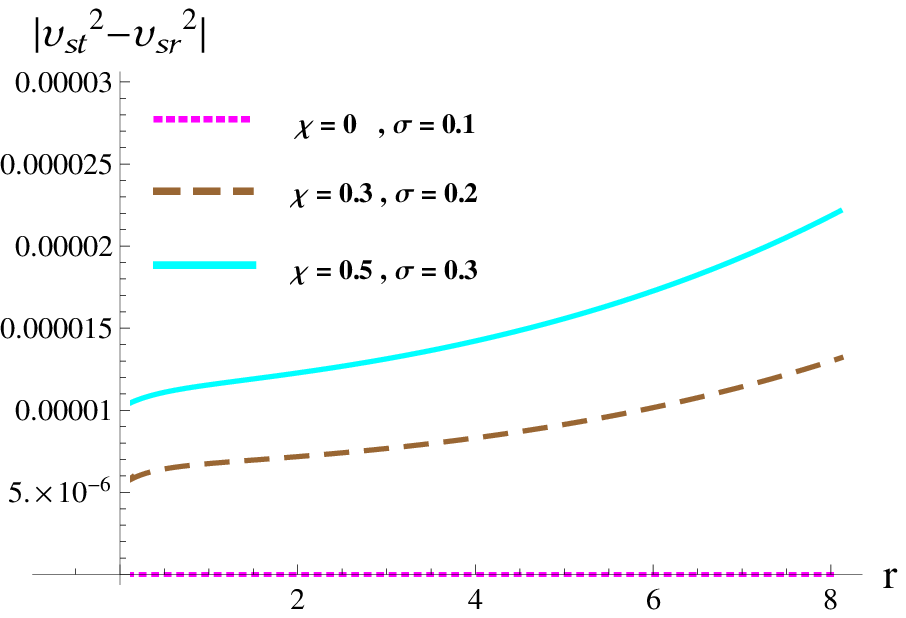,width=0.4\linewidth}\\
\caption{Plots of $v_{sr}^2$, $v_{st}^2$ and $|v_{st}^2-v_{sr}^2|$
for solution II.}\label{44d}
\end{figure}

\begin{figure}\center
\epsfig{file=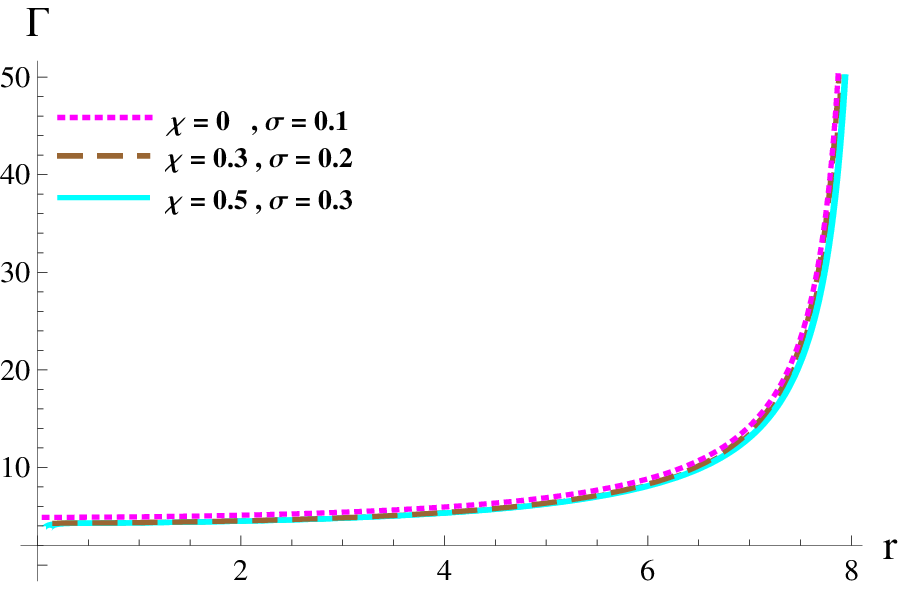,width=0.4\linewidth}\epsfig{file=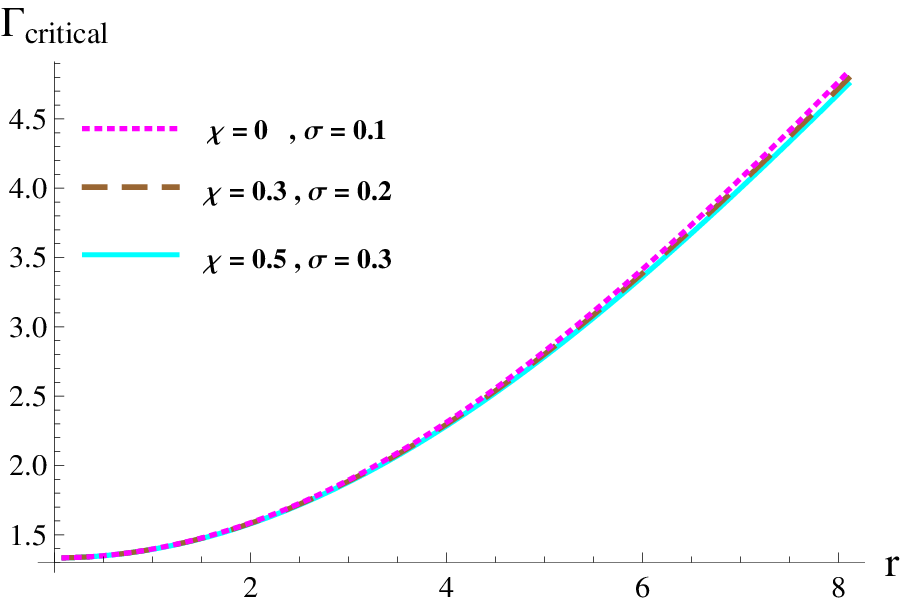,width=0.4\linewidth}\\
\caption{Plot of $\Gamma$ and $\Gamma_{critical}$ for solution
II.}\label{44e}
\end{figure}

In this solution, we choose three values of parameter $\chi=0,~0.3$
and $0.5$ and corresponding to $\sigma=0.1,~0.2$ and 0.3,
respectively. The profiles of $\widetilde{\rho}^{~eff}$,
$\widetilde{p}^{~eff}_{r}$, $\widetilde{p}^{~eff}_{t}$ and
$\widetilde{\Delta}^{eff}$ are displayed in Figure
\textbf{\ref{44a}}. It is observed that the energy density declines
and the radial/tangential pressure increases with the increasing
values of $\chi$ and $\sigma$. Moreover, matter variables are finite
within the interior of celestial object. The anisotropic factor
becomes zero at the center of object and increases for larger values
of $r$ and $\chi$. Moreover, for $\chi=0$, radial and tangential
pressures become equal leading to zero anisotropy. As all energy
conditions are satisfied therefore, the solution is physically
viable as shown in Figure \textbf{\ref{44b}}. From Figure
\textbf{\ref{44c}}, we can see that the system is in hydrostatic
equilibrium for the chosen values of $\chi$ and $\sigma$, as all the
three forces are balanced. Figures \textbf{\ref{44d}} and
\textbf{\ref{44e}} demonstrate the potential stability of the second
solution. The plots of radial and critical adiabatic index satisfy
the inequality $\Gamma\geq\Gamma_{critical}$ and lie above the
defined limit.

\section{Conclusions}

The formulation of new solutions for the study of self-gravitating
bodies has captured the interest of many astrophysicists. In this
regard, the EGD approach has effectively extended spherical
isotropic solutions by adding the anisotropic gravitational source.
In the current work, we have applied gravitational decoupling via
EGD to derive anisotropic solutions corresponding to the Starobinsky
model of $f(R)$ gravity. In order to check the consistency of the
EGD approach with this model, we have added the effects of a new
gravitational source in the isotropic Krori-Barua solution. The
$f(R)$ field equations for anisotropic fluid have been successfully
decoupled into two sets of equations with each array corresponding
to separate sources. The Bianchi identities for the matter sources
have indicated the transfer of energy between the two sources.
Furthermore, the constants in the considered solution have been
determined by the matching of interior and exterior spacetimes on
the boundary.

We have introduced a barotropic EoS for $\Theta_\eta^{\xi}$ as well
as imposed constraints on $\Theta_0^{0}$ and $\Theta_1^{1}$ which
has yielded solutions I and II, respectively. We have analyzed the
physical characteristics of the obtained solutions by plotting the
graphs of fluid parameters such as energy density, radial and
tangential pressures for the star Her X-I. It has been found that
the obtained solutions are physically well-behaved as they obey the
necessary conditions of viability. The energy density decreases with
a rise in values of $\chi$ which has led to the construction of less
dense spheres. On the other hand, the anisotropy attains larger
values for higher values of $\chi$ in both static solutions.
Moreover, the system corresponding to each solution is in
hydrostatic equilibrium. The proposed model corresponding to
solution I is potentially unstable according to the speed of sound
constraints while solution II is stable. We have also checked the
stiffness parameter corresponding to both solutions and found that
solution I violates the conditions $\Gamma>\frac{4}{3}$ and
$\Gamma\geq\Gamma_{critical}$ whereas solution II is consistent with
these criteria. Thus, the pressure of the developed spherical model
corresponding to solution II increases greatly in response to a
small change in density. Consequently, the compact body cannot be
compressed easily.

Sharif and Ama-Tul-Mughani \cite{26a} extended charged Krori-Barua
solution to the anisotropic domain via the EGD scheme in GR and
deduced that the solutions corresponding to both constraints are
physically viable and stable. The viability of the extended
solutions is preserved in $f(R)$ gravity while stability is
preserved for pressure-like constraint only. From the graphical
analysis, we have deduced that the second solution shows stable
behavior when $\chi\in[0,~0.5]$. Sharif and Waseem \cite{43} also
utilized the metric potentials of Krori-Barua solution to generate
anisotropic models by MGD approach in $f(R)$ gravity for $\chi$
ranging from $0$ to $1$. In comparison to this work, a smaller range
of $\chi$ generates stable anisotropic solutions corresponding to
the second constraint. It is worthwhile to mention here that the
$f(R)$ analog of this solution is physically viable and stable for
the particular values of the parameters $\sigma$ and $\chi$. We
conclude that the $f(R)$ theory yields stable decoupled stellar
configuration through EGD technique.

\section*{Appendix A}
\renewcommand{\theequation}{A\arabic{equation}}
\setcounter{equation}{0}

The modified terms appearing in the set related to the isotropic
source are
\begin{eqnarray}\nonumber
Y_{1}&=&4r^3\left(r\alpha'+4\right)\nu'\nu''+r^2\nu'^2\left(12r^2\alpha''
+3r^2\alpha'^2+16r\alpha'+16\right)+\nu\left(8\left(r^3\right.\right.\\\nonumber
&\times&\left.\left.\nu ^{(3)}\left(r\alpha'+4\right)+r^2\nu''
\left(4r^2\alpha''+r^2\alpha'^2+6r\alpha'(r)+4\right)-12\right)-2r
\nu'\left(r^3\right.\right.\\\nonumber
&\times&\left.\left.\alpha'^3+\alpha'\left(8r-18r^3\alpha''\right)-4
r^2\alpha'^2-8\left(3r^3\alpha^{(3)}+8r^2\alpha''-4\right)\right)\right)+\nu^2\\\nonumber
&\times&\left(-r^4\alpha'^4-8r^3\alpha'^3+16r^3\alpha'\left(r\alpha^{(3)}+\alpha
''\right)-4r^2\alpha'^2\left(r^2\alpha''+4\right)+4\right.\\\label{A1}
&\times&\left.\left(3r^4\alpha''^2+4\left(r^4\alpha^{(4)}+4r^3\alpha
^{(3)}+5\right)\right)\right)+16,\\\nonumber
Y_{2}&=&-r^2\left(r\alpha'+4\right)^2\nu'^2+2\nu\left(r^4\alpha'^3
\nu'+2r^3\alpha'^2\left(r\nu''+6\nu'\right)+4r^2\alpha'\left(\nu'\left(8\right.\right.\right.\\\nonumber
&+&\left.\left.\left.r^2\alpha''\right)+4r\nu''\right)+16\left(r^3\alpha''\nu'+2r^2
\nu''+3\right)\right)+\nu^2\left(-r^4\alpha'^4+4r^4\alpha'^2\right.\\\nonumber
&\times&\left.\alpha''+8r\alpha'\left(r^3\alpha^{(3)}+6r^2\alpha''-8\right)-4\left(r^4
\alpha''^2-8r^3\alpha^{(3)}-16r^2\alpha''\right.\right.\\\label{A2}
&+&\left.\left.28\right)\right)+16,\\\nonumber
Y_{3}&=&r^3\nu'^2\left(12r\alpha''+5r\alpha'^2+20\alpha'\right)+4r\nu'
\left(r^2\left(r\alpha'+4\right)\nu''+12\right)+2\nu\left(3r^4\right.\\\nonumber
&\times&\left.\alpha'^3\nu'+2r^3\alpha'^2\left(3r\nu''+8\nu'\right)+4r\alpha'
\left(r^3\nu^{(3)}+r\left(7r^2\alpha''-6\right)\nu'+7r^2\nu''\right.\right.\\\nonumber
&+&\left.\left.6\right)+8\left(2\left(r^4\alpha''\nu''+r^3\nu^{(3)}-3\right)+r\left(3r^3
\alpha^{(3)}+7r^2\alpha''-7\right)\nu'\right)\right)+\nu^2\\\nonumber
&\times&\left(r^4\alpha'^4+4r^3\alpha'^3+4r^2\alpha'^2\left(3r^2\alpha''-4\right)+8r\alpha'\left(3
r^3\alpha^{(3)}+5r^2\alpha''-2\right)\right.\\\label{A3}
&+&\left.4\left(4r^4\alpha^{(4)}+5r^4\alpha''^2+12r^3\alpha^{(3)}-8r^2\alpha''+28\right)\right)-16,
\end{eqnarray}
where the superscripts $(3)$ and $(4)$ indicate the third and fourth
order derivatives of the function with respect to $r$, respectively.
Equations (\ref{23})-(\ref{25}) contain the terms $Z_{1}$, $Z_{2}$
and $Z_{3}$, which are defined as
\begin{eqnarray}\nonumber
Z_{1}&=&\chi\left(r^4\chi^4g'^4+4r^3\chi^3\left(r\alpha'+2\right)g'^3+2r^2\chi^2\left(8
+3\alpha'^2r^2+2r^2\left(g''\chi\right.\right.\right.\\\nonumber
&+&\left.\left.\left.\alpha''\right)+12\alpha'r\right)g'^2+4r^2\chi\left(r^2\alpha'^3+6r\alpha'^2+2\left(\chi
g''r^2+\alpha''r^2+4\right)\alpha'\right.\right.\\\nonumber
&-&\left.\left.4r\left(\chi g''+\alpha''+r\chi
g^{(3)}+r\alpha^{(3)}\right)\right)g'+r^4\alpha'^4+8r^3\alpha'^3+4r^2\alpha'^2\left(\chi
g''r^2\right.\right.\\\nonumber
&+&\left.\left.\alpha''r^2+4\right)-16r^3\alpha'\left(\chi
g''+\alpha''+r \left(\chi
g^{(3)}+\alpha^{(3)}\right)\right)-4\left(3\chi^2 g''^2
r^4+3\right.\right.\\\nonumber
&\times&\left.\left.\alpha''^2r^4+6\chi g''\alpha''r^4+4\left(\chi
g^{(4)}r^4+\alpha^{(4)}r^4+4\chi
g^{(3)}r^3+4\alpha^{(3)}r^3+5\right)\right)\right)\\\nonumber
&\times&h^2+2\left(\chi^3g'^3\nu'r^4+\alpha'^3\nu'r^4+3\chi
g'\alpha'\nu'r^4\left(\alpha'+\chi
g'\right)-18\chi^2g'\nu'g''r^4\right.\\\nonumber
&-&\left.18\chi\alpha'\nu'g''r^4-4\chi
h''r^4\left(\chi^{2}g'^2+\alpha'^2\right)-8\chi^2g'\alpha'h''r^4-16\chi^2g''h''r^4\right.\\\nonumber
&-&\left.18\chi g'\nu'\alpha''r^4-18\alpha'\nu'\alpha''r^4-16\chi
h''\alpha''r^4-4\chi^2g'^2\nu''r^4-4\alpha'^2\nu''r^4\right.\\\nonumber
&-&\left.8\chi g'\alpha'\nu''r^4-16\chi g''\nu''r^4-16\alpha''\nu''
r^4-24\chi\nu' g^{(3)}r^4-4\chi^2g'h^{(3)}r^4\right.\\\nonumber
&-&\left.4\chi\alpha'h^{(3)} r^4-24\nu'\alpha^{(3)}r^4-4\chi g'\nu
^{(3)}r^4-4\alpha'\nu^{(3)}r^4-4\chi^2g'^2\nu'r^3\right.\\\nonumber
&-&\left.4\alpha'^2\nu'r^3-8\chi g'\alpha'\nu'r^3-64\chi\nu' g''r^3
-24\chi^2g'h''r^3-24\chi\alpha'h''r^3\right.\\\nonumber
&-&\left.64\nu'\alpha''r^3-24\nu''r^3\left(\chi g'+\alpha'\right)-16
\chi h^{(3)}r^3-16\nu^{(3)}r^3+8\chi g'\nu'r^2\right.\\\nonumber
&+&\left.8\alpha'\nu'r^2-16\chi h''r^2-16\nu''r^2+32\nu'r+\chi h'
\left(r^3\chi^3g'^3+r^2\chi^2g'^2\left(3r\alpha'\right.\right.\right.\\\nonumber
&-&\left.\left.\left.4\right)+r\chi\left(3\alpha'^2r^2-18r^2\left(\chi
g''+\alpha''\right)-8\alpha'r+8\right)g'+r^3\alpha'^3-4r^2\alpha'^2\right.\right.\\\nonumber
&-&\left.\left.2r\alpha'\left(9\chi g''r^2+9\alpha''r^2-4\right)-8
\left(-4+3\chi g^{(3)}r^3+3\alpha^{(3)}r^3+8\chi
g''r^2\right.\right.\right.\\\nonumber
&+&\left.\left.\left.8\alpha''r^2\right)\right)r+\nu\left(r^4\chi^4
g'^4+4r^3\chi^3\left(r\alpha'+2\right)g'^3+2r^2\chi^2\left(8+3\alpha'^2
r^2\right.\right.\right.\\\nonumber &+&\left.\left.\left.2\chi
g''r^2+2\alpha''r^2+12\alpha'r\right)g'^2+4r^2\chi\left(r^2\alpha'^3+6r
\alpha'^2+2\alpha'\left(\chi
g''r^2\right.\right.\right.\right.\\\nonumber
&+&\left.\left.\left.\left.\alpha''r^2+4\right)-4r\left(\chi
g''+\alpha''+r\chi g^{(3)}+r\alpha^{(3)}\right)\right)g'
+r^4\alpha'^4+8r^3\alpha'^3\right.\right.\\\nonumber
&+&\left.\left.4r^2\alpha'^2\left(\chi
g''r^2+\alpha''r^2+4\right)-16r^3\alpha'\left(\chi g''+\alpha''+r
\left(\chi
g^{(3)}+\alpha^{(3)}\right)\right)\right.\right.\\\nonumber
&-&\left.\left.4\left(3\chi^2 g''^2r^4+3\alpha''^2r^4+6\chi g''
\alpha''r^4+4\left(\chi g^{(4)}r^4+\alpha^{(4)}r^4+4\chi
g^{(3)}r^3\right.\right.\right.\right.\\\nonumber
&+&\left.\left.\left.\left.4\alpha^{(3)}r^3+5\right)\right)\right)+48\right)
h+r\left(r\left(r^2\chi^3g'^4+4r\chi^2\left(r\alpha'+2\right)g'^3+2\chi
g'^2\right.\right.\\\nonumber
&\times&\left.\left.\left(3\alpha'^2r^2+2\chi g''r^2+2\alpha''r^2
+12\alpha'r+8\right)+4\left(r^2\alpha'^3+6r\alpha'^2+2\alpha'\left(4\right.\right.\right.\right.\\\nonumber
&+&\left.\left.\left.\left.\chi g''r^2+\alpha''r^2\right)-4rg'\left(
\chi g''+\alpha''+r\chi g^{(3)}+r\alpha^{(3)}\right)\right)-4r
\left(-g''\alpha'^2r\right.\right.\right.\\\nonumber
&+&\left.\left.\left.4\left(g''+rg^{(3)}\right)\alpha'+3r\chi
g''^2+6rg''\alpha''+16g^{(3)}+4r g^{(4)}\right)\right)\nu^2+2
\left(h'\right.\right.\\\nonumber
&\times&\left.\left.\left(r^3\chi^3g'^3+r^2\chi^2\left(3r\alpha'-4\right)
g'^2+r\chi g'\left(3\alpha'^2r^2-18\chi
g''r^2-18\alpha''r^2\right.\right.\right.\right.\\\nonumber
&-&\left.\left.\left.\left.8\alpha'r+8\right)+r^3\alpha'^3-4r^2
\alpha'^2-2r\alpha'\left(9\chi g''r^2+9\alpha''r^2-4\right)-8
\left(-4\right.\right.\right.\right.\\\nonumber
&+&\left.\left.\left.\left.3\chi
g^{(3)}r^3+3\alpha^{(3)}r^3+8r^2\left(\chi
g''+\alpha''\right)\right)\right)+r\left(r^2\chi^2\nu'g'^3-r\chi
\left(\left(-3r\alpha'\right.\right.\right.\right.\right.\\\nonumber
&+&\left.\left.\left.\left.\left.4\right)\nu'+4r\left(\chi h''+\nu
''\right)\right)g'^2+\left(\nu'\left(3\alpha'^2r^2-18\chi g''r^2-18
\alpha''r^2-8\alpha'r\right.\right.\right.\right.\right.\\\nonumber
&+&\left.\left.\left.\left.\left.8\right)-4r\left(2\left(r\alpha'+3\right)\left(\chi
h''+\nu''\right)+r\left(\chi
h^{(3)}+\nu^{(3)}\right)\right)\right)g'-2\left(r\nu'\left(\left(9\right.\right.\right.\right.\right.\right.\\\nonumber
&\times&\left.\left.\left.\left.\left.\left.r\alpha'+32\right)
g''+12rg^{(3)}\right)+2\left(\left(4+\alpha'^2r^2+4r^2\left(\chi
g''+\alpha''\right)+6\alpha'r\right)\right.\right.\right.\right.\right.\\\nonumber
&\times&\left.\left.\left.\left.\left.h''+r\left(4r
g''\nu''+\left(r\alpha'+4\right)h^{(3)}\right)\right)\right)\right)\right)
\nu-r\left(\chi\left(3\chi^2g'^2r^2+3\alpha'^2r^2\right.\right.\right.\\\nonumber
&+&\left.\left.\left.12\chi g''r^2+12\alpha''r^2+16\alpha'r+2\chi g'
\left(3r\alpha'+8\right)r+16\right)h'^2+2\left(\nu'\left(16\right.\right.\right.\right.\\\nonumber
&+&\left.\left.\left.\left.3\chi^2g'^2r^2+3\alpha'^2r^2+12r^2\left(\chi
g''+\alpha''\right)+16\alpha'r+2\chi g'\left(3r\alpha'+8\right)
r\right)\right.\right.\right.\\\nonumber
&+&\left.\left.\left.2r\left(r\chi g'+r\alpha'+4\right)\left(\chi
h''+\nu''\right)\right)h'+r\nu'\left(3r\chi\nu'g'^2+2\left(\left(3r
\alpha'+8\right)\right.\right.\right.\right.\\\label{A4}
&\times&\left.\left.\left.\left.\nu'+2r\left(\chi
h''+\nu''\right)\right)g'+4
\left(3r\nu'g''+\left(r\alpha'+4\right)h''\right)\right)\right)\right),\\\nonumber
Z_{2}&=&\chi\left(\chi^4g'^4r^4+\alpha'^4r^4+4\chi^3g'^3\alpha'r^4-4
\alpha'^2\left(\chi
g''+\alpha''\right)r^4-2\chi^2g'^2\left(2\right.\right.\\\nonumber
&\times&\left.\left.\left(\chi g''+\alpha''\right)-3\alpha'^2\right)
r^4-8\alpha'r\left(\chi g^{(3)}r^3+\alpha^{(3)}r^3+6\chi g''r^2+6
\alpha''r^2\right.\right.\\\nonumber &-&\left.\left.8\right)-4\chi
g'\left(-\alpha'^3r^3+2\alpha'\left(\chi
g''+\alpha''\right)r^3+2\left(\chi
g^{(3)}r^3+\alpha^{(3)}r^3+6r^2\right.\right.\right.\\\nonumber
&\times&\left.\left.\left.\left(\alpha''+\chi
g''\right)-8\right)\right)r+4\left(\chi^2
g''^2r^4+\alpha''^2r^4-8r^3\left(\chi
g^{(3)}+\alpha^{(3)}\right)-16\right.\right.\\\nonumber
&\times&\left.\left.\alpha''r^2+2\chi
g''\left(r^2\alpha''-8\right)r^2+28\right)\right)h^2-2\left(\alpha'^3r^4\left(\chi
h'+\nu'\right)+\chi^3g'^3\right.\\\nonumber &\times&\left.\left(\chi
h'+\nu'\right)r^4-4\chi\alpha'g''r^4(\chi
h'+\nu')-2\chi\alpha'^2h''r^4-4\chi
h'\alpha'\alpha''r^4\right.\\\nonumber
&-&\left.4\alpha'\nu'\alpha''r^4-2\alpha'^2\nu''r^4-12\alpha
'^2r^3\left(\chi h'+\nu'\right)-16\chi g''r^3\left(\chi
h'+\nu'\right)\right.\\\nonumber &-&\left.16\chi\alpha'h''r^3-16\chi
h'\alpha''r^3-16\nu'\alpha''r^3-16\alpha'\nu''r^3-\chi^2g'^2\left(3
\chi h'\left(r\alpha'\right.\right.\right.\\\nonumber
&+&\left.\left.\left.4\right)+3\nu'\left(r\alpha'+4\right)+2r
\left(\chi h''+\nu''\right)\right)r^3-32r^2\left(\chi
h'\alpha'+\alpha'\nu'+\chi h''\right.\right.\\\nonumber
&+&\left.\left.\nu''\right)-\chi
g'\left(\nu'\left(r^2\left(3\alpha'^2+4\chi
g''+4\alpha''\right)+24\alpha'r+32\right)+\chi h'\left(3r^2
\alpha'^2\right.\right.\right.\\\nonumber
&+&\left.\left.\left.24r\alpha'+4\left(\chi
g''r^2+\alpha''r^2+8\right)\right)+4r\left(r\alpha'+4\right)
\left(\chi h''+\nu''\right)\right)r^2+\nu\right.\\\nonumber
&\times&\left.\left(\chi^4g'^4r^4+\alpha'^4r^4+4\chi^3g'^3\alpha'r^4-4\alpha'^2\left(\chi
g''+\alpha''\right)r^4-2\chi^2g'^2\left(2\left(\chi
g''\right.\right.\right.\right.\\\nonumber
&+&\left.\left.\left.\left.\alpha''\right)-3\alpha'^2\right)r^4-8
\alpha'\left(\chi g^{(3)}r^3+\alpha^{(3)}r^3+6\chi
g''r^2+6\alpha''r^2-8\right)r\right.\right.\\\nonumber
&-&\left.\left.4\chi g'\left(-\alpha'^3r^3+2\alpha'\left(\chi
g''+\alpha''\right)r^3+2\left(\chi
g^{(3)}r^3+\alpha^{(3)}r^3+6r^2\left(\chi
g''\right.\right.\right.\right.\right.\\\nonumber
&+&\left.\left.\left.\left.\left.\alpha''\right)-8\right)\right)r+4
\left(\chi^2g''^2r^4+\alpha''^2r^4-8\chi g^{(3)}r^3-8\alpha^{(3)}
r^3-16\alpha''r^2\right.\right.\right.\\\nonumber
&+&\left.\left.\left.2\chi g''\left(r^2\alpha''-8\right)r^2
+28\right)\right)-48\right)h+r\left(\left(r^3\chi^3g'^4+4r^3\chi^2
\alpha'g'^3\right.\right.\\\nonumber
&-&\left.\left.2r^3\chi\left(2\left(\chi g''+\alpha''\right)-3
\alpha'^2\right)g'^2+4\left(\alpha'^3r^3-2\alpha'\left(\chi
g''+\alpha''\right)r^3-2\right.\right.\right.\\\nonumber
&\times&\left.\left.\left.\left(\chi
g^{(3)}r^3+\alpha^{(3)}r^3+6\chi
g''r^2+6\alpha''r^2-8\right)\right)g'+4r\left(r^2\chi
g''^2-\left(\alpha'^2r^2\right.\right.\right.\right.\\\nonumber
&-&\left.\left.\left.\left.2\alpha''r^2+12\alpha'r+16\right)g''-2r
\left(r\alpha'+4\right)g^{(3)}\right)\right)\nu^2-2r\left(r^2\chi^2
\left(\chi h'\right.\right.\right.\\\nonumber
&+&\left.\left.\left.\nu'\right)g'^3+r\chi\left(3\chi
h'\left(r\alpha'+4\right)+3\nu'\left(r\alpha'+4\right)+2r\left(\chi
h''+\nu''\right)\right)g'^2\right.\right.\\\nonumber
&+&\left.\left.\left(\nu'\left(3\alpha'^2r^2+4\chi
g''r^2+4\alpha''r^2+24\alpha'r+32\right)+\chi h'\left(3r^2\alpha
'^2+24r\alpha'\right.\right.\right.\right.\\\nonumber
&+&\left.\left.\left.\left.4\left(\chi g''r^2+\alpha''r^2
+8\right)\right)+4r\left(r\alpha'+4\right)\left(\chi h''+\nu''
\right)\right)g'+\left(r\alpha'+4\right)\right.\right.\\\nonumber
&\times&\left.\left.\left(4r\nu'g''+2\left(r\alpha'+4\right)
h''+h'\left(r\alpha'^2+8\alpha'+4r\left(\chi
g''+\alpha''\right)\right)\right)\right)\nu+r\right.\\\nonumber
&\times&\left.\left(\chi h'^2\left(r\chi
g'+r\alpha'+4\right)^2+2h'\nu'\left(r\chi g'+r\alpha'+4\right)^2+r
g'\nu'^2\left(r\chi g'\right.\right.\right.\\\label{A5}
&+&\left.\left.\left.2r\alpha'+8\right)\right)\right),\\\nonumber
Z_{3}&=&\chi\left(r^4\chi^4g'^4+4r^3\chi^3\left(r\alpha'+1\right)
g'^3+2r^2\chi^2\left(3\alpha'^2r^2+6\chi
g''r^2+6\alpha''r^2\right.\right.\\\nonumber
&+&\left.\left.6\alpha'r-8\right)g'^2+4r\chi
\left(r^3\alpha'^3+3r^2\alpha'^2+\left(6r^3(\chi g''+\alpha'')-8
r\right)\alpha'+2\right.\right.\\\nonumber
&\times&\left.\left.\left(3r^3(\chi g^{(3)}+\alpha^{(3)})+5r^2(\chi
g''+\alpha'')-2\right)\right)g'+r^4\alpha'^4+4r^3\alpha'^3+4\right.\\\nonumber
&\times&\left.r^2\alpha'^2\left(3\chi g''r^2+3\alpha''r^2-4\right)+8
r\alpha'\left(3\chi g^{(3)}r^3+3\alpha^{(3)}r^3+5\chi
g''r^2\right.\right.\\\nonumber
&+&\left.\left.5\alpha''r^2-2\right)+4\left(5\chi^2g''^2r^4+5\alpha''^2r^4-8\alpha
''r^2+2\chi
g''\left(5r^2\alpha''-4\right)r^2\right.\right.\\\nonumber
&+&\left.\left.4\left(\chi g^{(4)}r^4+\alpha^{(4)}r^4+3\chi
g^{(3)}r^3+3\alpha^{(3)}r^3+7\right)\right)\right)h^2+2\left(3\chi
h'\alpha'^3r^4\right.\\\nonumber
&+&\left.3\alpha'^3\nu'r^4+3\chi^3g'^3\left(\chi h'+\nu'\right)
r^4+28\chi\alpha'g''r^4\left(\chi h'+\nu'\right)+6\chi\alpha'^2
h''r^4\right.\\\nonumber &+&\left.16\chi^2g''h''r^4+28\chi
h'\alpha'\alpha''r^4+28\alpha'\nu'\alpha''r^4+16\chi h''\alpha''
r^4+6\alpha'^2\nu''r^4\right.\\\nonumber &+&\left.16\chi
g''\nu''r^4+16\alpha''\nu''r^4+24\chi^2h'g^{(3)}r^4+24\chi\nu'
g^{(3)}r^4+4\chi\alpha'h^{(3)}r^4\right.\\\nonumber
&+&\left.24\alpha^{(3)}r^4(\chi
h'+\nu')+4\alpha'\nu^{(3)}r^4+16\alpha'^2r^3\left(\chi
h'+\nu'\right)+56\chi\left(\chi h'\right.\right.\\\nonumber
&+&\left.\left.\nu'\right)g''r^3+28\chi\alpha'h''r^3+56\chi h'
\alpha''r^3+56\nu'\alpha''r^3+28\alpha'\nu''r^3+\chi^2g'^2\right.\\\nonumber
&\times&\left.\left(\chi h'\left(9r\alpha'+16\right)+\nu'\left(9r
\alpha'+16\right)+6r\left(\chi
h''+\nu''\right)\right)r^3+16r^3\left(\chi
h^{(3)}\right.\right.\\\nonumber
&+&\left.\left.\nu^{(3)}\right)-24\chi h'\alpha'r^2-24\alpha'\nu'
r^2-56\chi h'r+24\alpha'r-56\nu'r+\chi g'\left(r\chi
h'\right.\right.\\\nonumber
&\times&\left.\left.\left(9r^2\alpha'^2+32r\alpha'+4\left(7\chi
g''r^2+7\alpha''r^2-6\right)\right)+r\nu'\left(9r^2\alpha'^2+32r
\alpha'\right.\right.\right.\\\nonumber
&+&\left.\left.\left.4\left(7\chi
g''r^2+7\alpha''r^2-6\right)\right)+4\left(\chi h^{(3)}r^3+\nu
^{(3)}r^3+\chi\left(3r\alpha'+7\right)h''r^2\right.\right.\right.\\\nonumber
&+&\left.\left.\left.\left(3r\alpha'+7\right)
\nu''r^2+6\right)\right)r+\nu\left(r^4\chi^4g'^4+4r^3\chi^3\left(r
\alpha'+1\right)g'^3+2r^2\chi^2\right.\right.\\\nonumber
&\times&\left.\left.\left(3\alpha'^2r^2+6\chi g''r^2+6\alpha''r^2+6
\alpha'r-8\right)g'^2+4r\chi\left(r^3\alpha'^3+3r^2\alpha'^2\right.\right.\right.\\\nonumber
&+&\left.\left.\left.\alpha'\left(6\chi
g''r^3+6\alpha''r^3-8r\right)+2\left(3r^3(\chi
g^{(3)}+\alpha^{(3)})+5\chi
g''r^2+5\alpha''r^2\right.\right.\right.\right.\\\nonumber
&-&\left.\left.\left.\left.2\right)\right)g'+r^4\alpha'^4+4r^3
\alpha'^3+4r^2\alpha'^2\left(3\chi g''r^2+3\alpha''r^2-4\right)+8r
\alpha'\left(-2\right.\right.\right.\\\nonumber
&+&\left.\left.\left.3\chi g^{(3)}r^3+3\alpha^{(3)}r^3+5\chi g''
r^2+5\alpha''r^2\right)+4\left(5r^4\left(\chi^2g''^2+\alpha''^2\right)\right.\right.\right.\\\nonumber
&-&\left.\left.\left.8\alpha''r^2+2\chi
g''\left(5r^2\alpha''-4\right)r^2+4\left(\chi g^{(4)}r^4+\alpha
^{(4)}r^4+3\chi g^{(3)}r^3\right.\right.\right.\right.\\\nonumber
&+&\left.\left.\left.\left.3\alpha^{(3)}r^3+7\right)\right)\right)-48\right)h+r
\left(\chi h'^2\left(5\chi g'(r\chi g'+2\left(r\alpha'+2\right))+5r
\alpha'^2\right.\right.\\\nonumber
&+&\left.\left.20\alpha'+12r\left(\chi g''+\alpha''\right)\right)
r^2+\nu'\left(5r\chi\nu'g'^2+2\left(5\left(r\alpha'+2\right)\nu'+2r
\left(\nu''\right.\right.\right.\right.\\\nonumber
&+&\left.\left.\left.\left.\chi h''\right)\right)g'+4\left(3r\nu'
g''+\left(r\alpha'+4\right)h''\right)\right)r^2+2h'\left(5\chi^2
g'^2\nu'r^3+5\alpha'^2\nu'r^3\right.\right.\\\nonumber
&+&\left.\left.2\alpha'\left(10\nu'+r\left(\chi
h''+\nu''\right)\right)r^2+2\chi g'\left(5\left(r\alpha'+2\right)
\nu'+r\left(\chi h''+\nu''\right)\right)r^2\right.\right.\\\nonumber
&+&\left.\left.4\left(3\nu'\left(\chi g''+\alpha''\right)r^3+2
\left(\chi h''r^2+\nu''r^2+3\right)\right)\right)+2\nu\left(3r^3
\chi^2g'^3\left(\chi h'\right.\right.\right.\\\nonumber
&+&\left.\left.\left.\nu'\right)+r^2\chi\left(\chi h'\left(9r\alpha
'+16\right)+\nu'\left(9r\alpha'+16\right)+6r\left(\chi
h''+\nu''\right)\right)g'^2\right.\right.\\\nonumber
&+&\left.\left.\left(r\chi h'\left(9r^2\alpha'^2+32r\alpha'+4
\left(7\chi g''r^2+7\alpha''r^2-6\right)\right)+r\nu'\left(9r^2
\alpha'^2\right.\right.\right.\right.\\\nonumber
&+&\left.\left.\left.\left.32r\alpha'+4\left(7\chi g''r^2+7\alpha
''r^2-6\right)\right)+4\left(\chi h^{(3)}r^3+\nu^{(3)}r^3+\chi
\left(3r\alpha'\right.\right.\right.\right.\right.\\\nonumber
&+&\left.\left.\left.\left.\left.7\right)h''r^2+\left(3r\alpha'+7\right)\nu''
r^2+6\right)\right)g'+2r^2\left(3rh''\alpha'^2+2\alpha'\left(7h''\right.\right.\right.\right.\\\nonumber
&+&\left.\left.\left.\left.rh^{(3)}\right)+2\nu'\left(7\left(r\alpha'+2\right)
g''+6rg^{(3)}\right)+8\left(rh''\alpha''+r g''\left(\chi
h''+\nu''\right)\right.\right.\right.\right.\\\nonumber
&+&\left.\left.\left.\left.h^{(3)}\right)\right)+h'\left(3r^3\alpha'^3+16
r^2\alpha'^2+4r\left(7\chi g''r^2+7\alpha''r^2-6\right)\alpha'+8
\left(3r^3\right.\right.\right.\right.\\\nonumber
&\times&\left.\left.\left.\left.\left(\chi
g^{(3)}+\alpha^{(3)}\right)+7\chi g''r^2+7\alpha''r^2-7\right)
\right)\right)+\nu^2\left(r^3\chi^3g'^4+4r^2\chi^2\left(r\alpha'\right.\right.\right.\\\nonumber
&+&\left.\left.\left.1\right)g'^3+2r\chi\left(3\alpha'^2r^2+6\chi
g''r^2+6\alpha''r^2+6\alpha'r-8\right)g'^2+4\left(r^3\alpha'^3\right.\right.\right.\\\nonumber
&+&\left.\left.\left.3r^2\alpha'^2+\left(6\chi g''r^3+6\alpha''r^3-8
r\right)\alpha'+2\left(3r^3(\chi
g^{(3)}+\alpha^{(3)})+5r^2\left(\alpha''\right.\right.\right.\right.\right.\\\nonumber
&+&\left.\left.\left.\left.\left.\chi
g''\right)-2\right)\right)g'+4r\left(5 r^2\chi
g''^2+\left(3\alpha'^2r^2+10\alpha''r^2+10\alpha'r-8\right)
g''\right.\right.\right.\\\label{A6}
&+&\left.\left.\left.2r\left(3\left(r\alpha'+2\right)g^{(3)}+2r
g^{(4)}\right)\right)\right)\right).
\end{eqnarray}
\section*{Appendix B}
\renewcommand{\theequation}{B\arabic{equation}}
\setcounter{equation}{0} The uncharged isotropic Krori-Barua
solution is obtained by adopting the following procedure.
Substituting expressions of $Y_{2}$ and $Y_{3}$ in Eqs.(\ref{20})
and (\ref{21}), respectively, we obtain
\begin{eqnarray}\nonumber
p_{r}&=&{\nu}\left(\frac{\alpha'}{r}
+\frac{1}{r^{2}}\right)-\frac{1}{r^{2}}+\frac{1}{8r^{4}}\sigma
\left[-r^2 \left(r\alpha'+4\right)^2\nu'^2+2\nu\left(r^4\alpha'^3
\nu'+2r^3\alpha'^2\right.\right.\\\nonumber
&\times&\left.\left.\left(r\nu''+6\nu'\right)+4r^2\alpha'\left(r^2
\alpha''\nu'+8\nu'+4r\nu''\right)+16\left(r^3\alpha''\nu'+2r^2\nu''\right.\right.\right.\\\nonumber
&+&\left.\left.\left.3\right)\right)+\nu^2\left(-\alpha'^4r^4+4r^4\alpha'^2\alpha''+8
r\alpha'\left(r^3\alpha^{(3)}+6r^2\alpha''-8\right)-4\left(r^4\alpha''^2\right.\right.\right.\\\label{A7}
&-&\left.\left.\left.8r^2\left(r\alpha^{(3)}+2\alpha''\right)+28\right)\right)+16\right],\\\nonumber
p_{t}&=&{\nu}\left(\frac{\alpha''}{2}
+\frac{\alpha'}{2r}+\frac{\alpha^{'2}}{4}\right)+\frac{\nu'}{2r}
+\frac{\nu'\alpha'}{4}+\frac{1}{8r^{4}}\sigma\left[r^3\nu'^2\left(12r
\alpha''+5r\alpha'^2\right.\right.\\\nonumber
&+&\left.\left.20\alpha'\right)+4r\nu'\left(r^2\left(r\alpha'+4\right)
\nu''+12\right)+2\nu\left(3r^4\alpha'^3\nu'+2r^3\alpha'^2\left(3r\nu
''\right.\right.\right.\\\nonumber
&+&\left.\left.\left.8\nu'\right)+4r\alpha'\left(r^3\nu^{(3)}+r
\left(7r^2\alpha''-6\right)\nu'+7r^2\nu''+6\right)+8\left(2
\left(r^4\alpha''\nu''\right.\right.\right.\right.\\\nonumber
&+&\left.\left.\left.\left.r^3\nu^{(3)}-3\right)+r\left(3r^3\alpha
^{(3)}+7r^2\alpha''-7\right)\nu'\right)\right)+\nu^2\left(r^4\alpha
'^4+4r^3\alpha'^3\right.\right.\\\nonumber
&+&\left.\left.4r^2\alpha'^2\left(3r^2\alpha''-4\right)+8r\alpha
'\left(3r^3\alpha^{(3)}+5r^2\alpha''-2\right)+4\left(4r^4\alpha
^{(4)}\right.\right.\right.\\\label{A8}
&+&\left.\left.\left.5r^4\alpha''^2+12r^3\alpha^{(3)}-8r^2\alpha
''+28\right)\right)-16\right].
\end{eqnarray}
Equations (\ref{A7}) and (\ref{A8}) are the $f(R)$ field equations
corresponding to radial and tangential pressures, respectively. The
fluid represents isotropic configuration if $p_{r}=p_{t}=p$.
Employing this condition yields
\begin{eqnarray}\nonumber
p&=&\frac{1}{8r^3}\left[2r\sigma\nu'^2\left(3r^2\alpha''+r^2\alpha
'^2+3r\alpha'-4\right)+\nu'\left(r^3\alpha'\left(2\sigma\nu''+1\right)+2\right.\right.\\\nonumber
&\times&\left.\left.\left(4r^2\sigma\nu''+r^2+12\sigma\right)\right)+\nu
\left(4r^3\sigma\alpha'^3\nu'+r^2\alpha'^2\left(8r\sigma\nu''+28\sigma
\nu'+r\right)\right.\right.\\\nonumber
&+&\left.\left.2\left(r\left(r^2\alpha''\left(8\sigma\nu''+1\right)+8r
\sigma\nu^{(3)}+16\sigma\nu''+2\right)+4\sigma\nu'\left(3r^3\alpha^{(3)}-7\right.\right.\right.\right.\\\nonumber
&+&\left.\left.\left.\left.9
r^2\alpha''\right)\right)+\alpha'\left(4r^3\sigma\nu^{(3)}
+8r\sigma\left(4r^2\alpha''+1\right)\nu'+44r^2\sigma\nu''+6r^2\right.\right.\right.\\\nonumber
&+&\left.\left.\left.24
\sigma\right)\right)+2\sigma\nu^2\left(r^2\alpha'^3+4r\alpha'^2\left(r^2\alpha''
-1\right)+4r\left(r^2\alpha^{(4)}+r^2\alpha''^2+2\alpha''\right.\right.\right.\\\label{A9}
&+&\left.\left.\left.5r\alpha^{(3)}\right)+\alpha'\left(8r^3\alpha^{(3)}+22r^2\alpha''
-20\right)\right)-4r\right],
\end{eqnarray}
which represents the expression of isotropic pressure. By plugging
the metric potentials of Krori-Barua solution, Eq.(\ref{A9}) turns
out to be
\begin{eqnarray}\nonumber
p&=&\frac{e^{-2\mathcal{A}r^2}}{2r^2}\left[-e^{2\mathcal{A}r^2}
+e^{\mathcal{A}r^2}\left\{1+\mathcal{B}^2r^4-\mathcal{A}\left(\mathcal{B}r^4
+r^2+12\sigma\right)+4\mathcal{B}\left(r^2\right.\right.\right.\\\nonumber
&+&\left.\left.\left.3\sigma\right)\right\}-4\sigma\left\{6
\mathcal{A}^3r^4\left(\mathcal{B}r^2+2\right)+\mathcal{A}\left(4\mathcal{B}^3
r^6+34\mathcal{B}^2r^4+35\mathcal{B}r^2-3\right)\right.\right.\\\label{A10}
&-&\left.\left.\mathcal{A}^2 r^2 \left(10 \mathcal{B}^2 r^4+43
\mathcal{B} r^2+20\right)+\mathcal{B} \left(-5 \mathcal{B}^2 r^4-11
\mathcal{B} r^2+3\right)\right\}\right].
\end{eqnarray}

\end{document}